\documentclass[preprint]{revtex4}

\usepackage{graphicx}

\begin{document}

\title{Phase competition in trisected superconducting dome}

\keywords{cuprates | photoemission | strongly correlated electron systems}

\author{I. M. Vishik}
\affiliation{Stanford Institute for Materials and Energy Sciences, SLAC National Accelerator Laboratory, 2575 Sand Hill Road, Menlo Park, CA 94025, USA}
\affiliation{Geballe Laboratory for Advanced Materials, Departments of Physics and Applied Physics, Stanford University, Stanford, CA 94305, USA}
\author{M Hashimoto}
\affiliation{Stanford Synchrotron Radiation Lightsource, SLAC National Accelerator Laboratory, 2575 Sand Hill Road, Menlo Park, CA 94025, USA}
\author {R.-H. He}
\affiliation{Advanced Light Source, Lawrence Berkeley National Lab, Berkeley, CA 94720, USA}
\author{W. S. Lee}
\affiliation{Stanford Institute for Materials and Energy Sciences, SLAC National Accelerator Laboratory, 2575 Sand Hill Road, Menlo Park, CA 94025, USA}
\affiliation{Geballe Laboratory for Advanced Materials, Departments of Physics and Applied Physics, Stanford University, Stanford, CA 94305, USA}
\author{F. Schmitt}
\affiliation{Stanford Institute for Materials and Energy Sciences, SLAC National Accelerator Laboratory, 2575 Sand Hill Road, Menlo Park, CA 94025, USA}
\affiliation{Geballe Laboratory for Advanced Materials, Departments of Physics and Applied Physics, Stanford University, Stanford, CA 94305, USA}
\author{D. H. Lu}
\affiliation{Stanford Synchrotron Radiation Lightsource, SLAC National Accelerator Laboratory, 2575 Sand Hill Road, Menlo Park, CA 94025, USA}
\author{R. G. Moore}
\affiliation{Stanford Institute for Materials and Energy Sciences, SLAC National Accelerator Laboratory, 2575 Sand Hill Road, Menlo Park, CA 94025, USA}
\author{C. Zhang}
\affiliation{State Key Laboratory of Crystal Materials, Shandong University, Jinan, 250100, P.R.China}
\author{W. Meevasana}
\affiliation{School of Physics, Suranaree University of Technology, Muang, Nakhon Ratchasima, 30000, Thailand}
\author{T. Sasagawa}
\affiliation {Materials and Structures Laboratory, Tokyo Institute of Technology, Meguro-ku, Tokyo 152-8550, Japan}
\author{S. Uchida}
\affiliation{Department of Physics, Graduate School of Science, University of Tokyo, Bunkyo-ku, Tokyo 113-0033, Japan}
\author{K. Fujita}
\affiliation{Laboratory for Atomic and Solid State Physics, Department of Physics, Cornell University, Ithaca, NY 14853, USA}
\author{S. Ishida}
\affiliation{Department of Physics, Graduate School of Science, University of Tokyo, Bunkyo-ku, Tokyo 113-0033, Japan}
\author{M. Ishikado}
\affiliation{Japan Atomic Energy Agency, Tokai, Ibaraki 319-1195, Japan}
\author{Y. Yoshida}
\affiliation{Superconducting Electronics Group, Electronics and Photonics Research Institute, National Institute of Advanced Industrial Science and Technology  (AIST), Ibaraki 305-8568, Japan}
\author{H. Eisaki}
\affiliation{Superconducting Electronics Group, Electronics and Photonics Research Institute, National Institute of Advanced Industrial Science and Technology  (AIST), Ibaraki 305-8568, Japan}
\author{Z. Hussain}
\affiliation{Advanced Light Source, Lawrence Berkeley National Lab, Berkeley, CA 94720, USA}
\author{T. P. Devereaux}
\affiliation{Stanford Institute for Materials and Energy Sciences, SLAC National Accelerator Laboratory, 2575 Sand Hill Road, Menlo Park, CA 94025, USA}
\affiliation{Geballe Laboratory for Advanced Materials, Departments of Physics and Applied Physics, Stanford University, Stanford, CA 94305, USA}
\author{Z.-X. Shen}
\email[To whom correspondence should be addressed. E-mail: ]{zxshen@stanford.edu}
\affiliation{Stanford Institute for Materials and Energy Sciences, SLAC National Accelerator Laboratory, 2575 Sand Hill Road, Menlo Park, CA 94025, USA}
\affiliation{Geballe Laboratory for Advanced Materials, Departments of Physics and Applied Physics, Stanford University, Stanford, CA 94305, USA}

\begin{abstract}
A detailed phenomenology of low energy excitations is a crucial starting point for microscopic understanding of complex materials such as the cuprate high temperature superconductors. Because of its unique momentum-space discrimination, angle-resolved photoemission spectroscopy (ARPES) is ideally suited for this task in the cuprates where emergent phases, particularly superconductivity and the pseudogap, have anisotropic gap structure in momentum space.  We present a comprehensive doping-and-temperature dependence ARPES study of spectral gaps in Bi$_2$Sr$_2$CaCu$_2$O$_{8+\delta}$  (Bi-2212), covering much of the superconducting portion of the phase diagram.   In the ground state, abrupt changes in near-nodal gap phenomenology give spectroscopic evidence for two potential quantum critical points, p$=$0.19 for the pseudogap phase and p$=$0.076 for another competing phase. Temperature dependence reveals that the pseudogap is not static below T$_c$ and exists p$>$0.19 at higher temperatures. Our data imply a revised phase diagram which reconciles conflicting reports about the endpoint of the pseudogap in the literature, incorporates phase competition between the superconducting gap and pseudogap, and highlights distinct physics at the edge of the superconducting dome.
\end{abstract}

\maketitle

The momentum-resolved nature of ARPES makes it a key probe of the cuprates whose interesting phases have anisotropic momentum-space structure \cite{Shen:dwave_first,Loeser:PG_science96,Ding:dwave_gap,Ding:PG}: both the
\textit{d}-wave superconducting gap and the pseudogap above T$_c$ have a maximum at the antinode (AN, near ($\pi$,0)) and are ungapped at the node, though the latter phase also exhibits an extended ungapped arc \cite{Marshall:FermiArcPocket1996,DestructionFS_Norman,Kanigel:FermiArc,Lee:twoGapARPES_TDep}. Ordering phenomena often result in gapping of the quasiparticle spectrum, and distinct quantum states produce spectral gaps with characteristic temperature, doping, and momentum dependence.  This was demonstrated by recent ARPES experiments that argued that the
pseudogap is a distinct phase from superconductivity based on their unique phenomenology \cite{Lee:twoGapARPES_TDep,Kaminski:2gapBi2201,HeHashimoto:Bi2201_eh_sym_break,Tanaka:twoGapARPES_dopingDep,Yoshida:LSCOGapFunction2009,KondoKaminski:TwoGapPRL2007,HeHashimoto:Science2011,Ma:CoexistanceCompetingOrdersBi2201}:  the pseudogap dominates near the antinode (AN) \cite{Tanaka:twoGapARPES_dopingDep,Lee:twoGapARPES_TDep}, and its magnitude increases with underdoping \cite{Tanaka:twoGapARPES_dopingDep,Yoshida:LSCOGapFunction2009}, whereas near-nodal (NN) gaps have a different doping dependence and can be attributed to superconductivity because they close at T$_c$ \cite{Lee:twoGapARPES_TDep,Yoshida:LSCOGapFunction2009}.  Previous measurements focused on AN or intermediate (IM) momenta, but laser-ARPES, with its superior resolution and enhanced statistics, allows for precise gap measurements near the node where the gap is smallest.  Our work is unique in its attention to NN momenta using laser-ARPES, and we demonstrate, via a single technique, that three distinct quantum phases manifest in different NN phenomenology as a function of doping.

\section{Results}
\begin{figure}[h]
\includegraphics [type=eps,ext=.eps,read=.eps,clip, width=3.5 in]{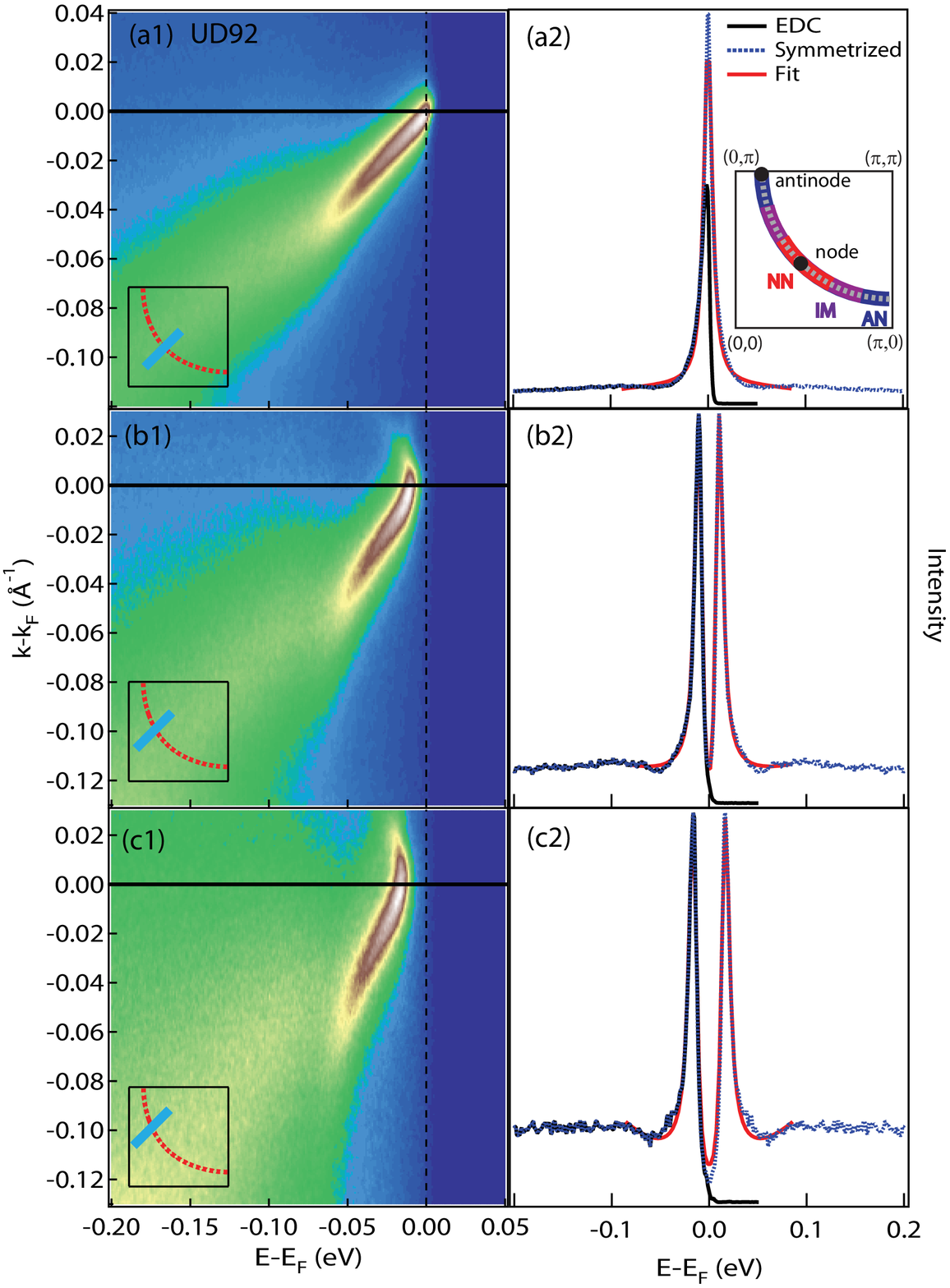}
\centering
\caption{\label{Fig 1:  Fig 1} Raw data, EDCs, and fitting.  (a1,b1,c1) Spectra for UD92 at the node (a1) and away from the node (b1,c1), with cut geometry and position shown in insets.  Horizontal lines are k$_F$. (a2,b2,c2) Raw (black) and symmetrized (blue) EDCs at k$_F$.  EDCs are fit to a minimal model \cite{Symmetrization_Norman_model} (red) to extract gap.  Inset of (a2) shows a quarter of the cuprate Brillouin zone and the FS (dotted line).  The node and antinode points are identified with black dots.  Locations of NN (red), IM (purple), and AN (blue) momenta are shown.}
\end{figure}

Gaps at parallel cuts were determined by fitting symmetrized energy distribution curves (EDCs) at k$_F$ to a minimal model \cite{Symmetrization_Norman_model}. The Fermi wavevector, k$_F$, is defined by the minimum gap locus. Example spectra, raw and symmetrized EDCs at k$_F$, and fits are shown for UD92 (underdoped, T$_c$$=$92) in Fig. \ref{Fig 1:  Fig 1}.

\subsection{Low Temperature}
\begin{figure*}[h]
\includegraphics [type=eps,ext=.eps,read=.eps,clip, width=6.5 in]{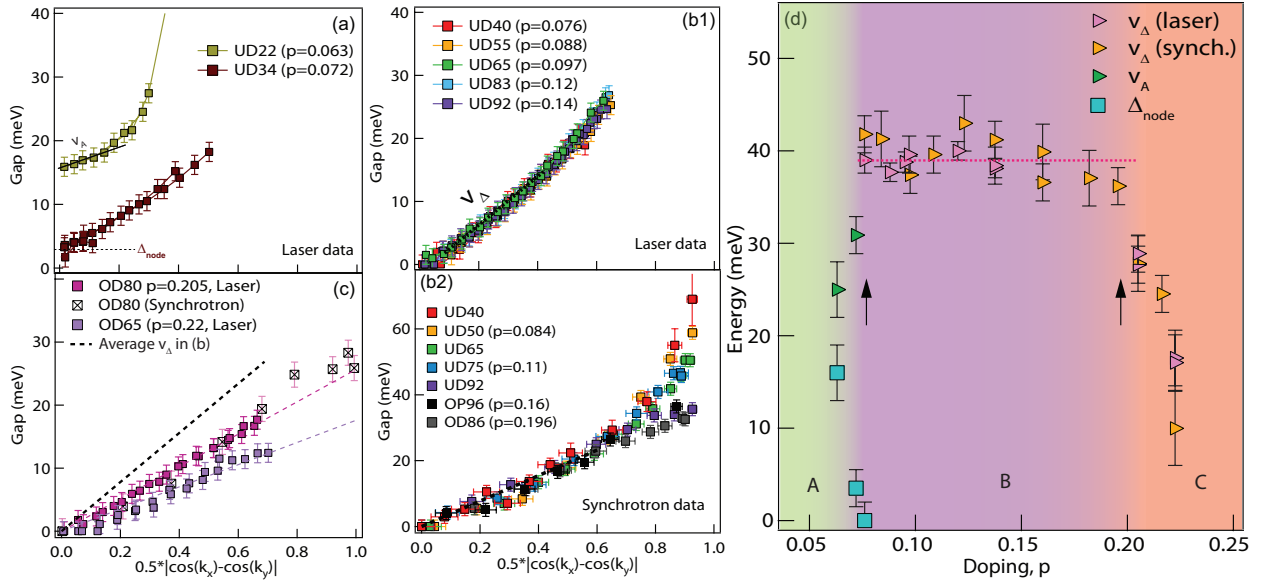}
\centering
\caption{\label{Fig 2:  Fig 2} Three distinct phase regions at low temperature.  Label UD/OP/OD denotes underdoped/optimal/overdoped sample with T$_c$ given by the number which follows.  (a)-(d) Gaps plotted in terms of the simple \textit{d}-wave form. v$_{\Delta}$ (v$_A$) is from a fit over the linear portion of the gap function, as shown by dotted(solid) line in (b1,a).  (a) In region \textbf{A}, FS is fully gapped with gap minimum, $\Delta$$_{node}$, at nodal momentum. Gap anisotropy v$_A$ decreases with underdoping. (b1-b2) Region \textbf{B} has doping-independent v$_{\Delta}$. (c) In region \textbf{C}, v$_{\Delta}$ decreases as T$_c$ decreases.  Dashed line is guide-to-the-eye for average v$_{\Delta}$ observed in region \textbf{B}.  Error bars in laser-ARPES reflect 3$\sigma$ error in fitting procedure and an additional 100$\%$ margin. Error bars in synchrotron data reflect uncertainty of determining E$_F$ ($\pm$0.5 meV), error from fitting procedure, and an additional 100$\%$ margin. (d) Summary of low-temperature NN energy scales. Arrows mark critical dopings defining three phase regions.}
\end{figure*}

Fig. \ref{Fig 2:  Fig 2}(a)-(c) shows gaps around the Fermi surface (FS) in terms of the simple \textit{d}-wave form, 0.5$|$$\cos$(k$_x$)$-$$\cos$(k$_y$)$|$, measured T$\approx$10K for the samples in our study (See SI Appendix for more details).  These data are quantified by the gap slope, v$_{\Delta}$, which measures how fast the \textit{d}-wave gap increases as a function of momentum away from the node.  We find that the low-temperature v$_{\Delta}$ changes suddenly at two dopings, p=0.076 and p=0.19, which are marked in the energy-doping phase diagram in Fig. \ref{Fig 2:  Fig 2}(d), dividing the superconducting dome into three regions, labeled \textbf{A}, \textbf{B}, and \textbf{C}.  In a \textit{d}-wave superconductor, v$_{\Delta}$ is expected to scale with T$_c$, and in region \textbf{C} (p$>$0.19, Fig. \ref{Fig 2:  Fig 2}(c)),  v$_{\Delta}$ and T$_c$  indeed decrease together.
Region \textbf{B} (0.076$\leq$p$\leq$0.19), exhibits a markedly different behavior: NN gaps are almost coincident over a large portion of the FS for  all samples shown in Fig. \ref{Fig 2:  Fig 2}(b1-b2), indicating a doping-\textit{independent} v$_{\Delta}$, despite T$_c$ varying more than two-fold.  The laser-ARPES gap functions show a slight curvature, nearly identical for all dopings, which is not visible in synchrotron-ARPES data, due to poorer resolution and a sparser sampling of momenta.  We note that the crossover between regions \textbf{B} and \textbf{C} is very abrupt, as v$_{\Delta}$ decreases by almost 25$\%$ for a change in doping $\Delta$p$=$0.01, after having been constant within error bars for $\Delta$p$=$0.12.
There is also a very abrupt transition between regions \textbf{A} and \textbf{B} at p=0.076 (Fig. \ref{Fig 2:  Fig 2}(a)), as region \textbf{A} is marked by a FS which is \textit{gapped at every momentum}.  The gap minimum ($\Delta$$_{node}$) is at the nodal momentum (along (0,0) - ($\pi$,$\pi$)) and increases with underdoping. Though v$_{\Delta}$ is no longer defined, the gap is still anisotropic around the FS.  We define a gap anisotropy parameter in region \textbf{A}, v$_A$, from the momentum dependence of the gap in Fig. \ref{Fig 2:  Fig 2}(a), and v$_A$  decreases with underdoping.  The low-temperature NN energy scales which characterize each of the three phase regions are summarized in Fig. \ref{Fig 2:  Fig 2}(d).   These findings are an important refinement to previous results which indicated that the NN region is dominated by superconductivity.  They demonstrate more conventional \textit{d}-wave superconductivity in region \textbf{C}, unconventional doping-independent \textit{d}-wave superconductivity in region \textbf{B}, and a nodeless unconventional superconductivity in region \textbf{A}.

\subsection{Temperature Dependence}
\begin{figure}[h]
\includegraphics [type=eps,ext=.eps,read=.eps,clip, width=3.5 in]{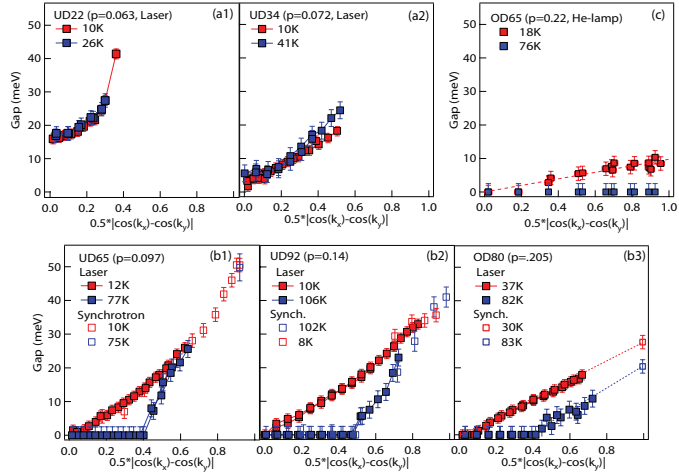}
\centering
\caption{\label{Fig 3:  Fig 3a} Distinct temperature dependence of gap in each of three phase regions. Red: low temperature gap.  Blue: gap T$>$T$_c$.  (a1-a2) In region \textbf{A}, NN gaps do not close across T$_c$.  (b1-b3) In region \textbf{B}, NN gaps partially close at T$_c$, with AN pseudogap remaining T$>$T$_c$.  OD80 is in phase region \textbf{C} at low temperature, but behaves like phase region \textbf{B} T$>$T$_c$.  (c) He-lamp data. For p$\geq$0.22, gap closes everywhere on FS  T$>$T$_c$.}
\end{figure}

In Fig. \ref{Fig 3:  Fig 3a} we compare low temperature gaps with gaps just above T$_c$ in each of the three phase regions.  For samples which are in region \textbf{A} at low temperature, the NN gaps are temperature-independent across T$_c$, and the FS remains fully gapped above T$_c$.  This indicates that the gap at the nodal momentum does not have a purely superconducting origin and that the onset doping for region \textbf{A} is the same at low temperature and T$_c$. In region \textbf{B}, gaps close or diminish near the node at T$_c$ while AN gaps remain above T$_c$. This observation of Fermi arcs near the node, defined as momenta where the symmetrized EDCs at k$_F$ are peaked at E$_F$ implying zero gap \cite{Kanigel:FermiArc}, and gaps near the antinode is the usual ARPES signature of the pseudogap above T$_c$.  All of the samples which exhibit characteristic doping-independent NN gaps of region \textbf{B} at low temperature also display a Fermi arc and antinodal gap above T$_c$.  Additionally, OD80 (p$\approx$0.205) has an AN gap persisting T$>$T$_c$, demonstrating a pseudogap above T$_c$ at this doping.  Thus, we classify the temperature-dependence of OD80 with region \textbf{B} in Fig. \ref{Fig 3:  Fig 3a} even though it exhibits region \textbf{C} phenomenology at low temperature.  This suggests that the doping separating regions \textbf{B} and \textbf{C} may be different at low temperature and at T$_c$ and that the pseudogap may exist at higher temperature for p$>$0.19.  The most overdoped sample in our study, OD65, is the only one to exhibit an ungapped FS T$>$T$_c$, demonstrating that the normal state pseudogap persists until p$\approx$0.22, in agreement with other recent ARPES results \cite{Chatterjee:PNAS_phase_diagramPG}.

\begin{figure}[h]
\includegraphics [type=eps,ext=.eps,read=.eps,clip, width=3.5 in]{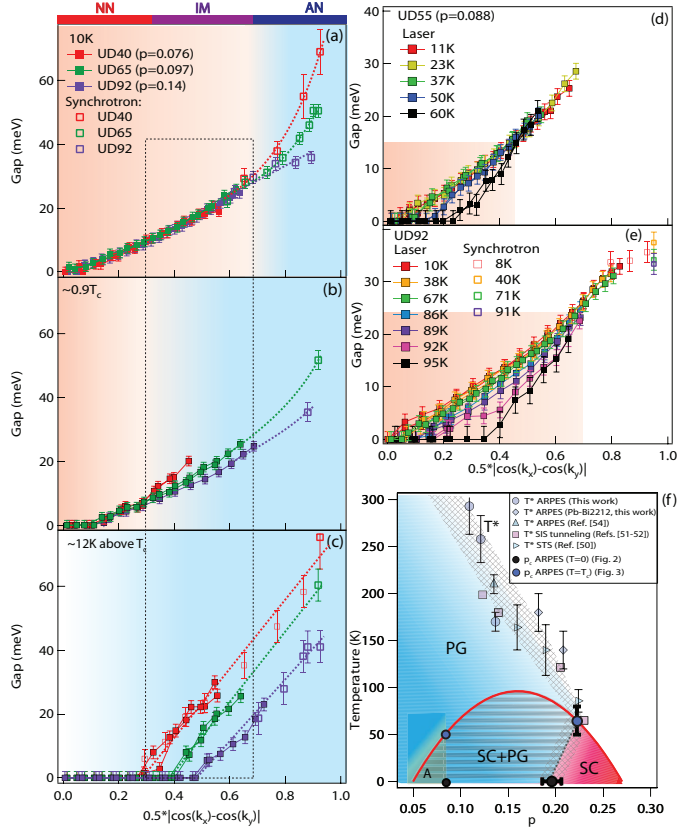}
\centering
\caption{\label{Fig 4:  Fig 4b} Phase competition in region \textbf{B} (a)-(c) Gaps in UD40, UD65, and UD92 at 10K, 0.9T$_c$, and 12K above T$_c$.  Synchrotron (laser) data shown with open (filled) symbols. Dashed lines are guides-to-the-eye. Doping-independent (dependent) gaps indicated by pink (blue) shading.  Dashed box marks momenta where gaps are doping-dependent in (b)-(c) but doping-independent in (a). (d)-(e) Gap functions at various temperatures below and across T$_c$ for UD55 and UD92 (see SI appendix for additional dopings).  Shaded regions denotes momenta where the gap T$>$T$_c$ is smaller than the low temperature gap.  Filled (open) symbols are laser(synchrotron)-ARPES data. (f) Proposed phase diagram.  T* is determined from ARPES measurements at antinode (SI appendix and Ref. \cite{Kondo:Fluct_vs_pseudogap}), STS \cite{Gomes:VisualizingPairFormation}, and SIS tunneling \cite{Dapisupil:JPSA_OD_PG_Bi2212_tunnel,Ozyuzer:PG_deeplyOD_Bi2212}.}
\end{figure}

We study temperature-and-doping dependence of gaps in two ways: doping dependence at comparable temperature and temperature dependence at varied dopings. The former is shown in Fig. \ref{Fig 4:  Fig 4b}(a)-(c), where three dopings (UD40, UD65, and UD92) are compared at three temperatures.  These dopings are chosen to be in a regime where the superconductivity and pseudogap energy scales are well separated in Bi-2212 (see SI appendix).  Two distinct doping dependencies are observed in different regions of the FS: doping-independent gaps and gaps which increase with underdoping.  At 10K, doping-independent gaps are observed at NN and IM momenta and gaps which increase with underdoping are observed at the AN.  Just below T$_c$, however, doping-dependent gaps extend into the IM region.  Above T$_c$, gaps increase with underdoping everywhere except the Fermi arc. Notably, a region of the FS, marked with a dashed box in Fig. \ref{Fig 4:  Fig 4b}(a)-(c) is home to doping-\textit{independent} gaps at low temperature but doping-\textit{dependent} gaps near/above T$_c$. Fig. \ref{Fig 4:  Fig 4b}(d)-(e) shows a full temperature dependence of gaps from low temperature to T$_c$ for UD55 and UD92.  Temperature dependence near the node occurs in a limited temperature range within $25\%$ of T$_c$, and the momentum region where gaps decrease near T$_c$ becomes larger with increasing doping.

\section{Discussion}
\subsection{Phase region A}
Whereas some ARPES experiments suggest a smooth evolution of phenomenology from the moderately underdoped regime to the edge of the superconducting dome \cite{Kanigel:FermiArc, Chatterjee:NatPhys_nodalLiquid, Johnson:SmallPocketDopingDepPRL}, our data indicate an emergent phase in region \textbf{A} (p$<$0.076) which coexists with superconductivity, characterized in ARPES by a gap at every FS momentum.  These results are supported by similar data in other cuprates.  A fully gapped state at the underdoped edge of the superconducting dome has been shown in Ca$_{2-x}$Na$_x$CuO$_2$Cl$_2$ (Na-CCOC) \cite{KM_shen:fully_gapped} and Bi$_2$Sr$_{2-x}$La$_x$CuO$_{6+\delta}$ (La-Bi2201) \cite{Hashimoto:NodalGap}, and our study is the first report of a fully gapped FS, both above and below T$_c$, in Bi-2212 at superconducting dopings. It is possible that region \textbf{A} represents an extension of pseudogap physics, but multiple spectroscopic changes p$<$0.076 together with reports of a distinct order at the underdoped edge of the superconducting dome in other compounds points to distinct physics.  There are three abrupt changes in NN gap phenomenology at p$=$0.076: a fully gapped FS appears, the gap anisotropy away from the nodal momentum starts to decrease, and the NN gaps become temperature-independent across T$_c$ such that Fermi arcs are not observed.  The latter result connects to in-plane transport in deeply underdoped cuprates which shows negative d$\rho$$_{ab}$$/$dT prior to the superconducting transition \cite{Ando:PhaseDiagramResistivity}. Notably, EDCs in region \textbf{A} remain sufficiently sharp near the nodal momentum (see SI appendix), such that it is unlikely that this behavior is primarily disorder driven. There have been recent reports of a similar critical doping p$\approx$0.07-0.10 in YBa$_2$Cu$_3$O$_y$ (YBCO), varyingly attributed to a metal-insulator quantum critical point \cite{Sebastian:Diverging_effective_mass_QO}, a Lifshitz transition \cite{LeBoeuf:LifshitzCriticalPoint_YBCO}, or spin density wave order (SDW) \cite{Haug:SDW_YBCO}.  Though a similar onset doping might suggest a common origin of phenomena observed in YBCO and Bi-2212, there are some inconsistencies, such as thermal conductivity data, which do not support a fully gapped FS at low dopings in YBCO \cite{Sutherland:ThermalMetal2005}.  This discrepancy may be materials-dependent, reflecting differences in disorder and Fermiology. Alternately, the ground state in region \textbf{A} may exhibit intrinsic time or spatial variation such that different techniques are sensitive to different aspects, which is supported by neutron scattering and muon spin-relaxation measurements in YBCO indicating slowly fluctuating spin order at the edge of the superconducting dome \cite{Hinkov:Neutron_mSR_YBCO_ELC}.  It has been shown that SDW order can gap nodal quasiparticles \cite{Nazario:SDW_dSC}, so this may be common to both compounds.

\subsection{Phase Regions B and C}
Though superconductivity has been shown to dominate at NN momenta \cite{Lee:twoGapARPES_TDep,Kaminski:2gapBi2201}, Fig. \ref{Fig 2:  Fig 2}(b) indicates that NN gaps are remarkably insensitive to T$_c$ in a broad doping range constituting region \textbf{B}, highlighting that NN gaps in region \textbf{B} do not reflect the bare superconducting order parameter. This doping-independent v$_{\Delta}$ is supported by specific heat measurements in YBCO \cite{Tallon:SF_density_cuprate_new_paradigm} and from scanning tunneling spectroscopy (STS) data in Bi-based cuprates \cite{Boyer:ImagingTwoGapsBi2201,Pushp:UniversalNode}.  Our data are the most complete ARPES demonstration of this behavior, crucially revealing p=0.076 and p=0.19 as dopings where region \textbf{B} abruptly ends.  The sudden change in v$_{\Delta}$ at p=0.19 is interpreted as the T$=$0 endpoint of the pseudogap.  This assignment has a precedent from low-temperature experiments which indicated that both the superfluid density and the Cu-site impurity-doping needed to suppress superconductivity are maximum at p$=$0.19 \cite{Anukool:LondonPenetration2009,Panagopoulos:SF_density_Hg1201_LSCO,Tallon:SF_density_cuprate_new_paradigm}.  Additionally, earlier ARPES data showed a maximum in the antinodal quasiparticle spectral weight, one measure of the strength of superconductivity relative to other spectral features, at p$=$0.19 \cite{Feng:SPR}.  In the interpretation that p$=$0.19 is the T$=$0 endpoint of the pseudogap, the more conventional relation between T$_c$ and v$_{\Delta}$ in region \textbf{C} reflects a pure superconducting ground state at low temperature, and region \textbf{B} is identified as a coexistence regime of superconductivity and the pseudogap.  Coexistence of superconductivity and pseudogap in region \textbf{B} has support both from other experiments and from independent ARPES data in our study.  STS experiments show symmetry-breaking order associated with the pseudogap throughout this doping range \cite{Parker:FluctStripePseudogapSTM}, and intrinsic tunneling spectroscopy shows distinct superconducting and pseudogap features below T$_c$ \cite{Krasnov:InstrinsicTunnelingTwoGap}. In the present study, for all of region \textbf{B}, coexistence of pseudogap and superconductivity in Bi-2212 manifests via distinct temperature dependence of gaps near the node and further away from the node \cite{Lee:twoGapARPES_TDep}. For the most underdoped portion of region \textbf{B} (p$<$0.12), coexistence also manifests in a gap function which deviates strongly from a simple \textit{d}-wave form at the AN, such that v$_{\Delta}$$<$$\Delta$$_{AN}$, where $\Delta$$_{AN}$ is the antinodal gap (see SI Appendix).  The doping where $\Delta$$_{AN}$ first surpasses v$_{\Delta}$ is not significant, and simply indicates the doping where the superconducting gap (NN) energy scale is sufficiently smaller than the pseudogap (AN) energy scale.  For some lower T$_c$ cuprates, gaps already deviate from a simple \textit{d}-wave form at optimal doping and show stronger deviation than Bi-2212 in the underdoped regime \cite{Yoshida:LSCOGapFunction2009,Kaminski:2gapBi2201,He:LBCO_one_eigth}. Although the pseudogap is considered to be primarily an antinodal phenomenon, our results demonstrate that it also manifests at NN momenta in region \textbf{B} via the doping-independent v$_{\Delta}$.  Similarly, the absence of the pseudogap in the ground state p$>$0.19 is also apparent at NN momenta, via a doping-dependent v$_{\Delta}$.  The remarkable doping-independent v$_{\Delta}$ in region \textbf{B} remains unexplained, but it may indicate a superconducting gap whose magnitude is renormalized by coexistence with the pseudogap.

The temperature dependence of spectral gaps provides microscopic information about the dynamics of the superconductivity/pseudogap coexistence in region \textbf{B}.  Fig. \ref{Fig 4:  Fig 4b}(a)-(c) demonstrates that at IM momenta, gaps have characteristic doping dependence of T*--increasing with underdoping \cite{DopingDependencePG:Tallon,Campuzano:PDH}-- when superconductivity is weak (absent) just below (above) T$_c$, but are doping-independent at low temperature.  This shows that the pseudogap is not static below T$_c$, but rather, it is suppressed by superconductivity at low temperature.  This nuance within the 'two-gap' picture indicates that the temperature dependence of the pseudogap must also be considered for quantitative understanding of the superconducting state.   Fig. \ref{Fig 4:  Fig 4b}(a)-(c) shows that the Fermi arc just above T$_c$ does not represent the only momenta where superconductivity emerges, because the doping-independent gap region at T$=$10K extends beyond the Fermi arc measured T$>$T$_c$.  A better way to define momenta with superconducting character is by temperature dependence near T$_c$, and the superconductivity-dominated momentum region defined in this manner expands with doping.  Although a pure superconducting gap closes entirely at T$_c$, we use a more lenient definition--a gap which diminishes approaching T$_c$--to define momenta with superconducting character.  We note that superconductivity exists over the entire FS in Bi-2212, as sharp quasiparticles are observed at the AN for p$>$0.08 \cite{Vishik:QPI_ARPES, Niestemski:VBG_theory_PG, Feng:SPR}, but our definition of the 'superconducting region' selects the portion of the FS where the temperature dependence of gaps indicates significant spectral contributions from superconductivity. This definition also permits for coexistence of pseudogap and superconductivity at some momenta, accounts for the observation that the pseudogap itself has temperature dependence, and is not hindered by difficulties in defining the Fermi arc length due to its temperature dependence \cite{Kanigel:FermiArc}.  The temperature dependence data in Fig. \ref{Fig 4:  Fig 4b}(a)-(e) provides a phase competition picture of superconductivity/pseudogap interaction in momentum space: the pseudogap is suppressed by superconductivity at low temperatures and larger dopings, and it surrenders a portion of the FS where it once existed.

\subsection{Proposed Phase Diagram}
The starting point of the phase diagram proposed in Fig. \ref{Fig 4:  Fig 4b}(f) is the observation of three distinct phase regions at low temperature as a function of doping, separated by two potential quantum critical points inside the superconducting dome at p=0.076 and p=0.019.  The former marks the onset of region \textbf{A}, possibly related to SDW order, while the latter is interpreted as the T=0 endpoint of the pseudogap.  Both critical points have support from other experiments in numerous cuprates \cite{Anukool:LondonPenetration2009,Panagopoulos:SF_density_Hg1201_LSCO,Tallon:SF_density_cuprate_new_paradigm,Zheng:Bi2201_pc_NMR,Sebastian:Diverging_effective_mass_QO, Haug:SDW_YBCO,KM_shen:fully_gapped}, and some theoretical proposals also favor a ground state with multiple critical points \cite{Chakravarty:QO_densityWave}. Our data demonstrate how these potential critical points manifest in the phenomenology of NN spectral gaps measured by ARPES.
The phase diagram in Fig. \ref{Fig 4:  Fig 4b}(f) also features conjectured re-entrant behavior of the pseudogap inside the superconducting dome, as a direct consequence of phase competition between superconductivity and the pseudogap \cite{Wu:CompetingOrders,Gabovich:Competition_SC_SDW,Ekino:PhaseDiagramCoexistanceCDW}.  The phase boundary between regions \textbf{B} (SC$+$PG) and \textbf{C} (SC) is anchored by ARPES data at T$=$0 and T$=$T$_c$, which show a sudden change in v$_{\Delta}$ and an absence of pseudogap T$>$T$_c$, respectively.  It is supported by OD80 data which obeys region \textbf{C} phenomenology at low temperature, but region \textbf{B} phenomenology at higher temperature, with the pseudogap persisting above T$_c$. It has been shown that the T$=$0 endpoint of a competing order is expected to shift under the superconducting dome \cite{Sachdev:QCP_shift}, such that high temperature measurements of the pseudogap phase boundary do not extrapolate to the T=0 endpoint seen inside the superconducting dome.  This manifests clearly in the BaFe$_2$As$_2$ family of iron pnictides compounds where both magnetic and structural phases have been shown to coexist with and be suppressed by superconductivity \cite{Wiesenmayer:MicroscopicCoexistanceMagnetismSCPnictide,Nandi:122PhaseDiagram}, and a phase diagram with re-entrant behavior has been demonstrated \cite{Nandi:122PhaseDiagram}.  A phase diagram with a reentrant pseudogap resolves conflicting reports about the fate of the pseudogap transition temperature, T*, inside the superconducting dome.  Some experiments suggest that the T* line intersects the superconducting dome and reaches T$=$0 at p=0.19 \cite{DopingDependencePG:Tallon,Tallon:SF_density_cuprate_new_paradigm}, whereas others, particularly spectroscopies, including our ARPES measurements of T* shown in Fig. \ref{Fig 4:  Fig 4b}(f), indicate that T* and T$_c$ merge on the strongly overdoped side \cite{Gomes:VisualizingPairFormation,Dapisupil:JPSA_OD_PG_Bi2212_tunnel,Ozyuzer:PG_deeplyOD_Bi2212,Chatterjee:PNAS_phase_diagramPG}.  Though variations between different experiments are expected, our data uniquely demonstrate both behaviors using a single technique.

\section{Conclusions}
We have performed a thorough doping-and-temperature dependence study of spectral gaps in superconducting Bi-2212.  At low temperature, we report three distinct phase regions with different characteristic phenomenology of NN gaps.  In phase region \textbf{B} (0.076$<$p$<$0.19), which is identified as a regime where superconductivity coexists with the pseudogap in the ground state, gaps at NN and IM momenta are independent of doping.  In region \textbf{C} (p$>$0.19), identified as a pure superconducting ground state, the \textit{d}-wave superconducting gap decreases as T$_c$ decreases.  Region \textbf{A} (p$<$0.076) is identified as an emergent phase characterized by a fully gapped FS and a gap anisotropy which decreases with underdoping.  Temperature dependence of gaps reveals phase competition between the pseudogap and superconductivity, where pseudogap physics dominates a smaller region of the FS at low temperatures and larger dopings.  From these doping-and-temperature dependence data we propose a new phase diagram featuring a trisected superconducting dome and re-entrant behavior of the pseudogap.

\section{Materials}
Lab-based experiments were done with 7 eV  laser or monochromated He-I light (21.2 eV) (Gammadata He lamp) and a Scienta SES2002 analyzer.  7 eV photons were produced by second harmonic generation from a 355 nm laser (Paladin, Coherent, Inc.) using a nonlinear crystal KBe$_2$BO$_3$F$_2$. Laser energy and momentum resolution were 3 meV and  better than 0.005 \AA$^{-1}$, respectively.  Synchrotron data were taken at the Stanford Synchrotron Radiation Lightsource with a Scienta R4000 analyzer and energy resolution approximately 8 meV. Samples were cleaved at 10-30K  \textit{in situ} at a pressure $<$4$\times$10$^{-11}$ torr to obtain a clean surface.  Doping was determined from T$_c$ via an empirical curve, T$_c$$=$T$_{c,max}$$*$[1-82.6(p-0.16)$^2$], taking 96K as the optimum T$_c$ for Bi-2212 \cite{UniversalCurve}.

\section{Acknowledgments}
We thank S. Kivelson, C. Varma, D. J. Scalapino, and S. Sachdev for helpful discussions.  This work is supported by the Department of Energy, Office of Basic Energy Science under Contract No. DEAC02-76SF00515.

\section{Note}
During the review process, following the submission of this manuscript, related papers appeared in support of charge density wave order, possibily related to the pseudogap, which competes with superconductivity \cite{Ghiringhelli:IC_charge_fluct_YBCO,Chang:Competition_SC_CDW_hard_xray}.  In addition, another paper reported a gap at the nodal momentum in LSCO \cite{Razzoli:NodalGapLSCO}.

\clearpage

\end{document}


\title{Phase competition in trisected superconducting dome: Supporting Information}
\maketitle

\renewcommand{\thesection}{S.\arabic{section}}
\renewcommand{\thesubsection}{\thesection.\arabic{subsection}}

\renewcommand{\bibnumfmt}[1]{[S#1]}
\renewcommand{\citenumfont}[1]{S#1}

\tableofcontents
\clearpage

\section{Samples}
\begin{table}[h]
\caption{Summary of samples shown in Fig. 2 of manuscript with their composition and experimental setup. $\Gamma$Y refers to cuts taken parallel to the (0,0)-($\pi$,$\pi$) line and $\Gamma$M refers to cuts taken parallel to ($\pi$,0)-($\pi$,$\pi$).  Dopings in manuscript determined from T$_c$ via an empirical curve, T$_c$$=$T$_{c,max}$$*$[1-82.6(p-0.16)$^2$], taking 96K as the optimum T$_c$ for Bi-2212 \cite{UniversalCurve}.}

\begin{tabular}{ l c c r }

Sample & Composition & Temperature (Fig. 2) & Experiment \\
  \hline
  UD22 & Bi$_2$Sr$_2$(Ca,Dy)Cu$_2$O$_{8+\delta}$ & 10 & 7eV, $\Gamma$Y \\
  UD34 & Bi$_2$Sr$_2$(Ca,Dy)Cu$_2$O$_{8+\delta}$ & 11 & 7eV, $\Gamma$Y \\
  UD40 & Bi$_2$Sr$_2$(Ca,Dy/Y)Cu$_2$O$_{8+\delta}$ & 12 & 7eV, $\Gamma$Y; 19eV, $\Gamma$Y \\
  UD50 & Bi$_2$Sr$_2$(Ca,Y)Cu$_2$O$_{8+\delta}$ & 10 & 19eV, $\Gamma$Y \\
  UD55 & Bi$_2$Sr$_2$(Ca,Dy)Cu$_2$O$_{8+\delta}$ & 11 & 7eV, $\Gamma$Y \\
  UD65 & Bi$_{2+x}$Sr$_{2-x}$CaCu$_2$O$_{8+\delta}$ & 12 & 7eV, $\Gamma$Y \\
  UD75 & Bi$_2$Sr$_2$CaCu$_2$O$_{8+\delta}$ & 10 & 22.7eV, $\Gamma$M \\
  UD83 & Bi$_2$Sr$_2$CaCu$_2$O$_{8+\delta}$ & 13 & 7eV, $\Gamma$Y \\
  UD85 & Bi$_2$Sr$_2$CaCu$_2$O$_{8+\delta}$ & 13 & 22.7eV, $\Gamma$M \\
  UD92 & Bi$_2$Sr$_2$CaCu$_2$O$_{8+\delta}$ & 10 & 7eV, $\Gamma$Y; 22.7eV, $\Gamma$M\\
  OP96 & Bi$_2$Sr$_2$(Ca,Y)Cu$_2$O$_{8+\delta}$ & 10 & 21.2eV, $\Gamma$Y \\
  OP98 & (Bi,Pb)$_2$Sr$_2$CaCu$_2$O$_{8+\delta}$ & 30 & 18.4eV, $\Gamma$M  \\
  OD92 & (Bi,Pb)$_2$Sr$_2$CaCu$_2$O$_{8+\delta}$ & 10 & 18.4eV, $\Gamma$Y  \\
  OD86 & Bi$_2$Sr$_2$CaCu$_2$O$_{8+\delta}$ & 18 & 22.7eV, $\Gamma$M  \\
  OD80 & (Bi,Pb)$_2$Sr$_2$CaCu$_2$O$_{8+\delta}$ & 12,30 & 7eV, $\Gamma$Y; 18.4eV, $\Gamma$M  \\
  OD71 & (Bi,Pb)$_2$Sr$_2$CaCu$_2$O$_{8+\delta}$ & 30 & 18.4eV,  $\Gamma$M  \\
  OD65 & (Bi,Pb)$_2$Sr$_2$CaCu$_2$O$_{8+\delta}$ & 10, 18 & 7eV, 21.2eV, $\Gamma$Y  \\
\end{tabular}
\label{Table 1}
\end{table}
 \clearpage

\section{Fitting}

\begin{figure*} [h]
\includegraphics [type=eps,ext=.eps,read=.eps,clip, width=6 in]{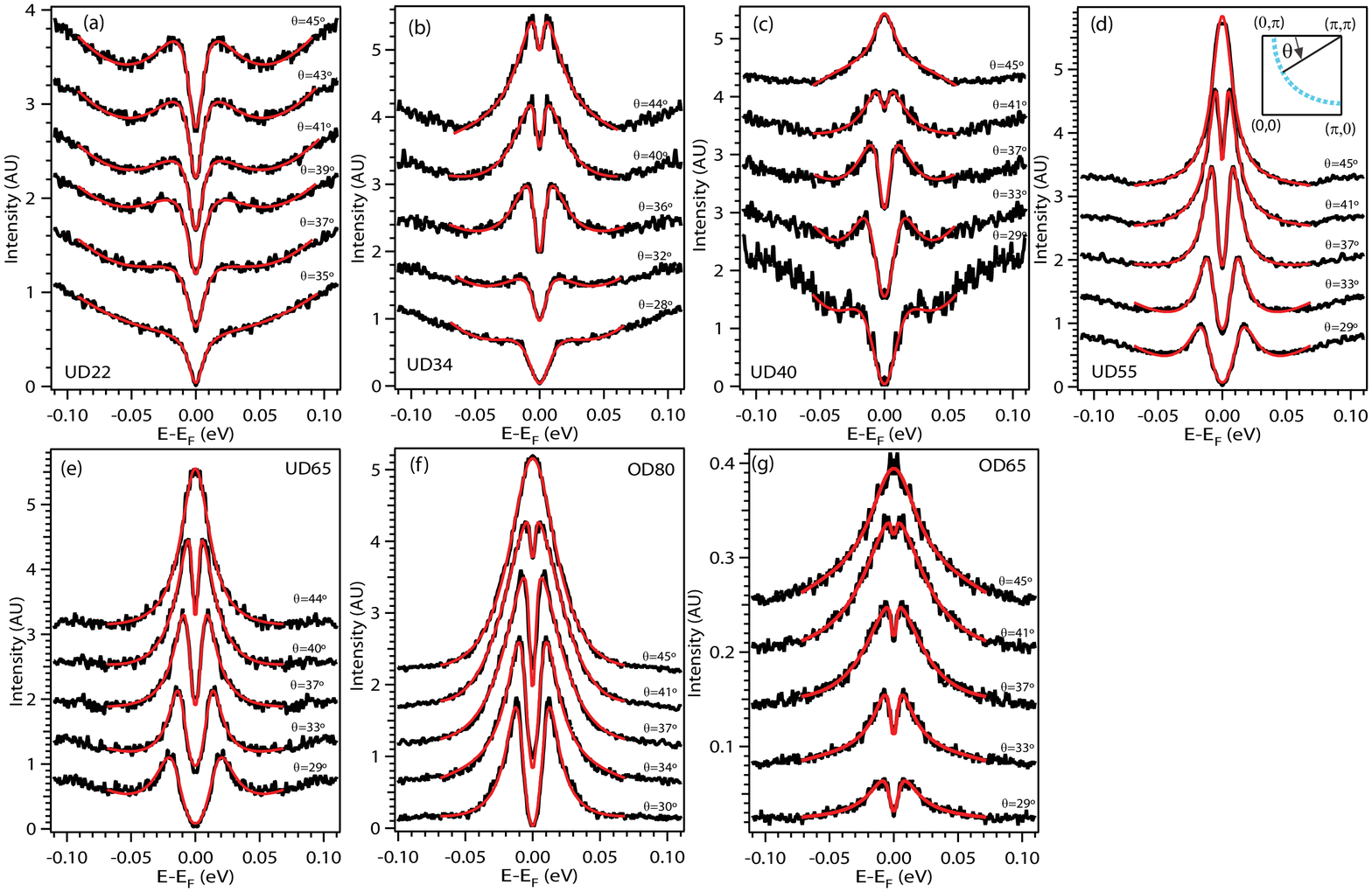}
\centering
\caption{\label{Fig 1: EDCs at low temperature} Selected symmetrized EDCs at low temperatures with fits.  All data taken with 7eV laser and cuts parallel to $\Gamma$Y.}
\end{figure*}

The energy gap in energy distribution curves (EDCs) can be quantified by several metrics: the position of the leading edge midpoint (LEM) relative to E$_F$, the energy positions of a quasiparticle peak, or by fitting data to an assumed model.  The first two methods do not take the lineshape into account, and are less suitable for comparing gaps among samples with different dopings. Thus, we determined the gap at each cut by fitting symmetrized EDCs at the Fermi wavevector, k$_F$, to a minimal model proposed by Norman \textit{et al.} \cite{Symmetrization_Norman_model}, $\Sigma$(\textbf{k}, $\omega$)=-\textit{i}$\Gamma$$_1$+ $\Delta$$^2$/[($\omega$+\textit{i}0$^+$) + $\epsilon$(\textbf{k})], where $\Gamma$$_1$ is a single particle scattering rate, $\epsilon$(\textbf{k}) is the dispersion, and the gap, $\Delta$, is the quantity of interest in the fitting.  It is assumed that $\epsilon$(\textbf{k$_F$})=0, and k$_F$ is defined by the minimum gap locus.  A quadratic background was also included to fully account for the lineshape in the deeply underdoped regime or at momenta far from the node. This fitting is applicable to our data as long as a peak is visible in the EDC. In Fig. S\ref{Fig 1: EDCs at low temperature}, we show the low energy portion ($\omega$$<$110meV) of symmetrized EDCs at low temperature together with fits.  In laser ARPES data, EDC peaks become smaller away from the node, which is not intrinsic for most dopings.  Synchrotron data taken at higher photoenergy with cuts parallel to $\Gamma$M do not show such a substantial decrease in peak intensity \cite{Vishik:QPI_ARPES,Vishik:NJP}, with the exception of deeply underdoped samples (p$<$0.09).  The intensity of the quasiparticle peak relative to the higher energy part of the spectrum is also generally not intrinsic, but comparisons between different dopings can be made if experimental conditions (photoenergy, polarization, cut geometry) are identical.  EDC peaks become smaller and broader with decreasing doping, a correlation and disorder effect, as widely reported \cite{Feng:SPR}, and the model we use provides a good fit to all data throughout the doping range, even though it is a minimal model and does not capture the full physics of the system.  In phase region \textbf{A}, the near-nodal EDCs show low-energy peaks which are narrow enough for a gap energy to be assessed accurately, though these peaks are not strictly quasiparticle-like because the width is larger than the binding energy.

\begin{figure*} [h]
\includegraphics [type=eps,ext=.eps,read=.eps,clip, width=7 in]{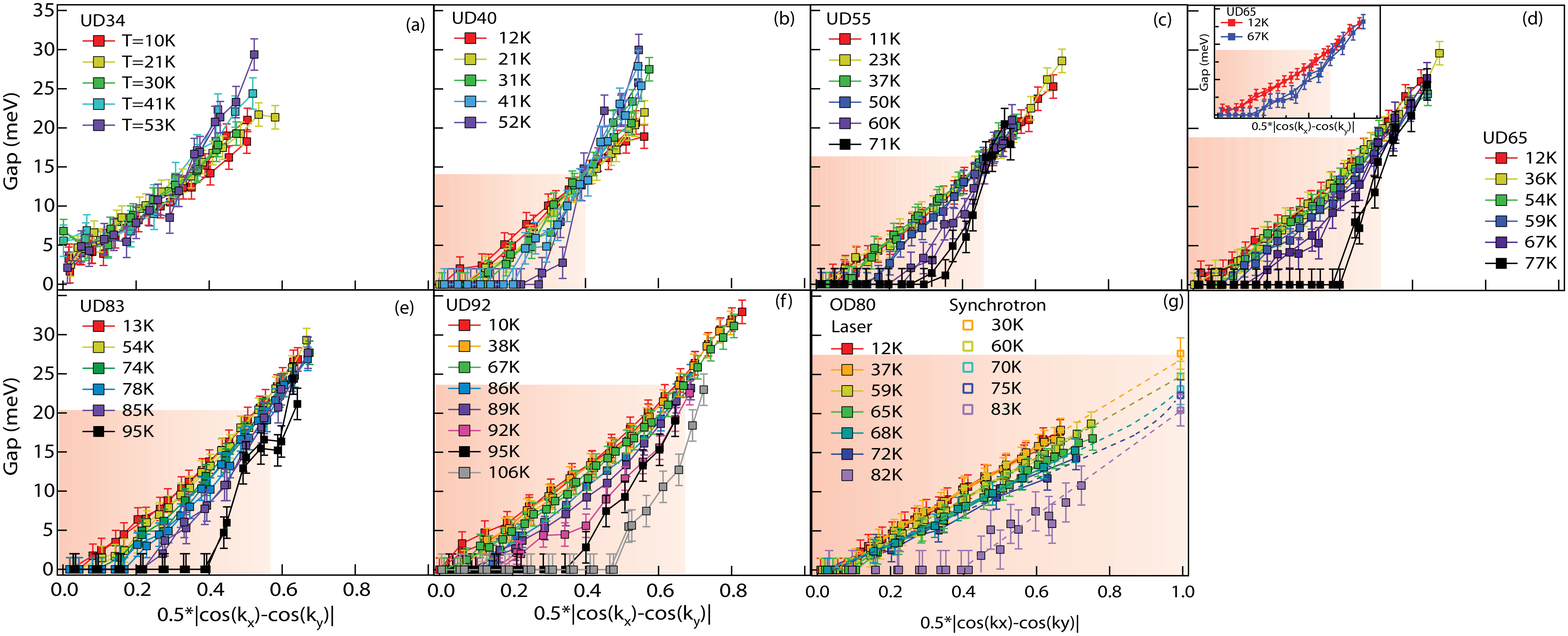}
\centering
\caption{\label{Fig 2: Full Temperature Dependence} Gaps from low temperature to T$>$T$_c$.  Shaded region denotes momenta where gap T$>$T$_c$ is smaller than low temperature gap, as explained in inset of (d).}
\end{figure*}

\begin{figure} [h]
\includegraphics [type=eps,ext=.eps,read=.eps,clip, width=3.5 in]{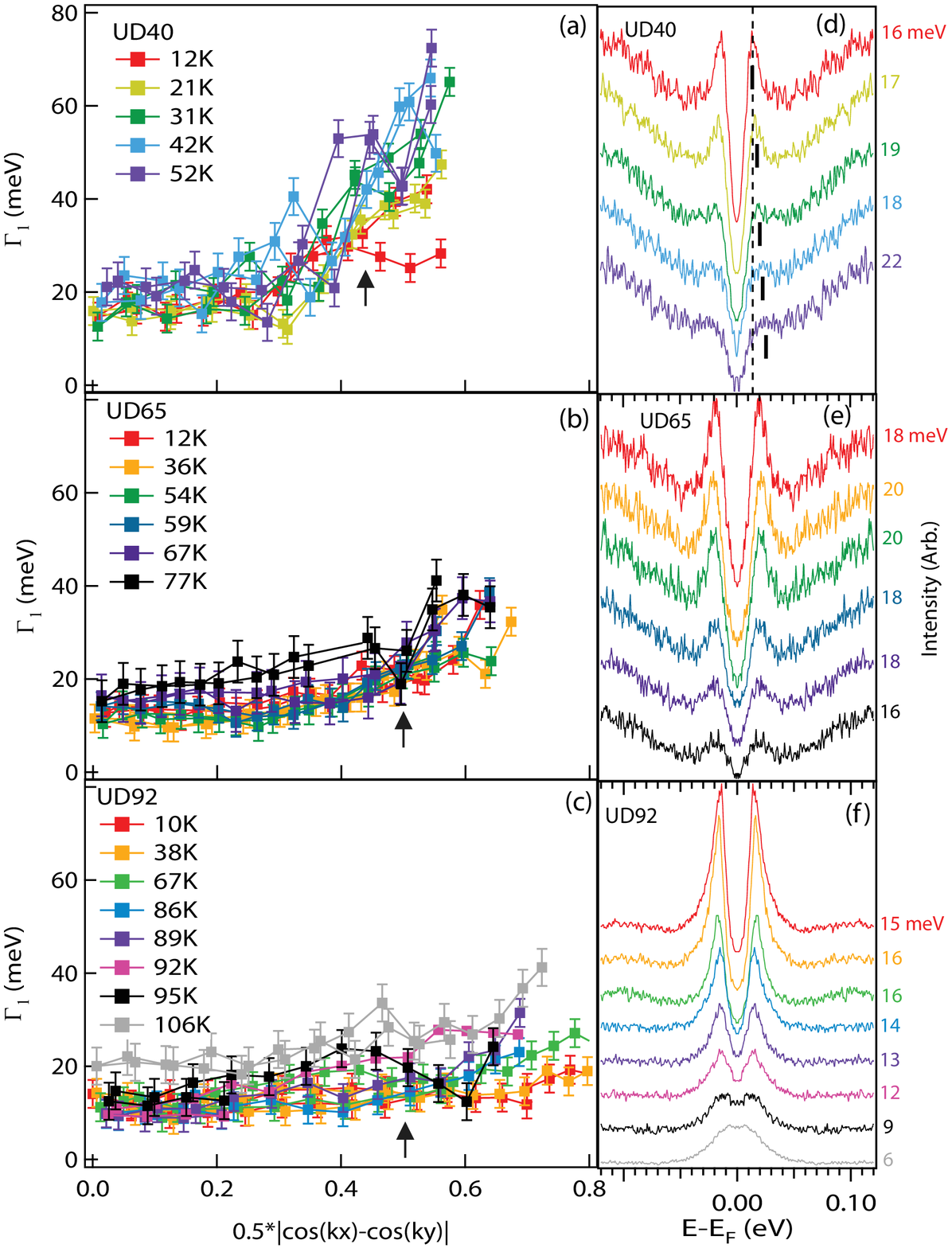}
\centering
\caption{\label{Fig 3:  Fitting gamma}  Scattering rates and temperature dependence of EDCs.  (a)-(c) $\Gamma$$_1$ \cite{Symmetrization_Norman_model} from fitting for selected dopings.  Error bars denote average 3$\sigma$ confidence interval for each temperature. (d)-(f) Symmetrized EDCs at the cut position indicated by arrow in (a)-(c).  Numbers to the right of panels indicate the gap value from fitting the EDC.  In panel (d), dashed line denotes EDC peak position at lowest temperature, while short vertical lines denote peak positions at all temperatures.}
\end{figure}

Figure S\ref{Fig 2: Full Temperature Dependence} shows gaps at all measured temperatures for a number of dopings, as a supplement to Fig. 4(d)-(e) in the manuscript.  The momentum region where the gap diminishes near T$_c$ is shaded in pink. Figure S\ref{Fig 3:  Fitting gamma} shows the single particle scattering rate $\Gamma$$_1$ from fitting for select samples (UD40, UD65, UD92), and EDCs at a selected momentum.  We note that there are momenta for UD40 and UD34 where the fitted gap increases slightly with temperature, and this is also visible in raw EDCs, as shown in Fig. S\ref{Fig 3:  Fitting gamma}(d).

\section{Antinodal gaps, comparison to previously published data, and scaling with T$_c$}

\begin{figure} [h]
\includegraphics [type=eps,ext=.eps,read=.eps,clip, width=3.5 in]{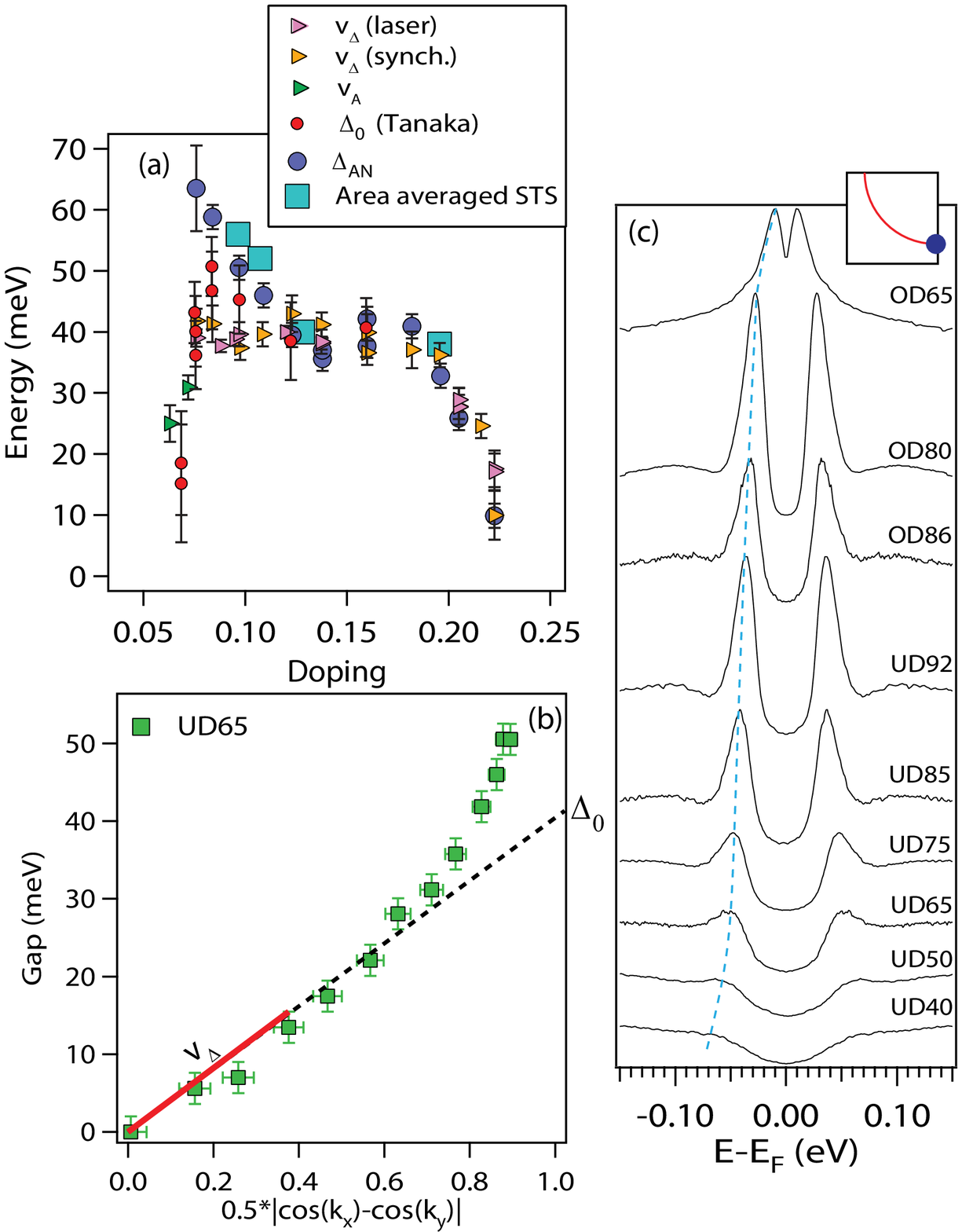}
\centering
\caption{\label{Fig 4:  Comparison to kiyo} Previously published data and antinodal gaps. (a) Comparison of near-nodal gap slope to previously published data (Tanaka, et al. Ref. \cite{Tanaka:twoGapARPES_dopingDep}).  Data are consistent where they overlap except new results draw decisive conclusions about doping independence of v$_{\Delta}$ for 0.076$\leq$p$\leq$0.19 and show that decrease of gap slope in deeply underdoped regime (v$_A$) happens in conjunction with the opening of a gap at the nodal momentum. Antinodal gaps ($\Delta$$_{AN}$) are determined from fitting symmetrized EDCs.   A gap function close to a simple \textit{d}-wave form is realized when $\Delta$$_{AN}$$\approx$v$_{\Delta}$.  Area-averaged STS energy gap from peak position at positive bias from curves in Ref. \cite{Lawler:NematicOrder}.  (b) Definition of $\Delta$$_0$ from Ref. \cite{Tanaka:twoGapARPES_dopingDep} and v$_{\Delta}$ from manuscript. (c) symmetrized EDCs at the antinode, in order of increasing doping from bottom to top.  Dotted line is guide-to-the-eye for peak position.}
\end{figure}

Tanaka \textit{et al.} previously reported the doping-dependence of the antinodal extrapolation of the near-nodal gap $\Delta$$_0$ \cite{Tanaka:twoGapARPES_dopingDep}, a quantity nominally equivalent to the near-nodal gap slope, v$_{\Delta}$.  A comparison between the data published by Tanaka \textit{et al.} and data in this manuscript are shown in Fig. S\ref{Fig 4:  Comparison to kiyo}.  The precision of laser ARPES allows us to draw the more definitive conclusion that near-nodal gaps are independent of doping for 0.076$\leq$p$\leq$0.19.  Fig. S\ref{Fig 4:  Comparison to kiyo} also shows $\Delta$$_{AN}$, the gap extracted from fitting symmetrized EDCs at the antinode.  All data are T$\ll$T$_c$.  When the gap function deviates strongly from a simple \textit{d}-wave form, ($\Delta$$_{AN}$$>($v$_{\Delta}$,$\Delta$$_0$)),  $\Delta$$_0$  will depend on how much of the near-nodal Fermi surface (FS) is considered in the extrapolation, which is why there is a larger difference between $\Delta$$_0$ and v$_{\Delta}$ for p$<$0.12.

$\Delta$$_{AN}$ is extracted by fitting the energy position of the superconducting quasiparticle peak at the antinode (the strongly peaked features in Fig. S\ref{Fig 4:  Comparison to kiyo}(c)), and it is plotted in Fig. S\ref{Fig 4:  Comparison to kiyo}(a).  Values quantitatively agree with area-averaged STS  \cite{Lawler:NematicOrder}.  $\Delta$$_{AN}$ increases with underdoping p$<$0.12 (pseudogap energy scale sufficiently dominates superconductivity), shows weak doping dependence for 0.12$\leq$p$\leq$0.19 (superconductivity and pseudogap have similar energy scales), and decrease with increasing doping p$>$0.19 (superconductivity over entire FS in ground state). While the energy position of the antinodal quasiparticle peak ($\Delta$$_{AN}$) can be strongly influenced by the underlying pseudogap, the distinction between the two is important.  This is clearly illustrated for the case of La-Bi2201 (Ref. \cite{HeHashimoto:Science2011}) where the energy scales of superconductivity and the pseudogap are well separated: the antinodal superconducting feature appears as a shoulder at 30 meV, the antinodal pseudogap feature appears as a broad hump near 70 meV, and the simple \textit{d}-wave extrapolation of near-nodal gaps to the antinode is 15 meV.  The energy position of the superconducting shoulder feature at the antinode is not that of near-nodal superconductivity or the antinodal pseudogap, but it is affected by both--a superconducting feature whose energy position is pushed to higher binding energy near the antinode because of the underlying pseudogap. Similarly, in Bi-2212 when the gap function deviates strongly from a simple \textit{d}-wave form near the antinode, the energy position of the antinodal quasiparticle peak ($\Delta$$_{AN}$) is not a measure of superconducting or pseudogap order parameters.  However, $\Delta$$_{AN}$ does follow the doping dependence of T* in the doping regime p$<$0.12, indicating that it reflects strong pseudogap physics.  In Fig. 4(a)-(c) of the manuscript, the dopings UD40, UD65, and UD92 are chosen to be in a doping regime where superconductivity and pseudogap energy scales are separated to varying degree.  The gaps plotted in those figures are always derived from superconducting features, but the differing doping and temperature dependencies arise from varying influences of the underlying pseudogap on the energy position of the superconducting features.

As discussed in the manuscript, the precise doping where the gap function deviates from a simple \textit{d}-wave form depends on the relative energy scales of the antinodal pseudogap and the near-nodal superconductivity.  Fig. S\ref{Fig 4:  Comparison to kiyo} indicates that there is a doping range 0.12$\leq$p$\leq$0.19 where $\Delta$$_{AN}$ is almost independent of doping and the gap function is close to a simple \textit{d}-wave form (defined at v$_{\Delta}$$\approx$$\Delta$$_{AN}$); notably, in this doping range, T* decreases with doping (Fig. 4(f) of manuscript).  This itself is a non-trivial observation which provides additional evidence that the pseudogap is suppressed by superconductivity below T$_c$, because the antinodal region assumes the doping-independence of near-nodal gaps, rather than the doping-dependence of T*.  It must be noted that although a slight curvature away from a simple \textit{d}-wave form is observed in laser-ARPES data for UD83 and UD92, both with p$\geq$0.12, but v$_{\Delta}$$\approx$$\Delta$$_{AN}$ in those samples, such that near-nodal and antinodal energy scales are similar and the gap function is not considered to deviate strongly from a simple \textit{d}-wave form.   Nevertheless, this slight curvature of the gap function may be important for understanding subtleties of pseudogap/superconductivity coexistence.

For p$<$0.076, our new data shows somewhat similar behavior to Ref. \cite{Tanaka:twoGapARPES_dopingDep}, in that the \textit{slope} of the near-nodal gaps decrease with further underdoping, but the interpretation is different because laser ARPES reveals a gap at the nodal momentum in region \textbf{A}.  In the simplest scenario, the gap measured below T$_c$ in region \textbf{A} represents a sum of a \textit{d}-wave superconducting gap ($\Delta$$_{SC}$(\textbf{k})), a momentum-independent gap ($\Delta$$_{node}$), and a momentum-dependent pseudogap ($\Delta$$_{PG}$(\textbf{k})) of the form $\Delta$$_A$$^2$$=$$\Delta$$_{SC}$(\textbf{k})$^2$$+$$\Delta$$_{node}$$^2$$+$$\Delta$$_{PG}$(\textbf{k})$^2$.  Thus, v$_A$ may indeed reflect \textit{d}-wave superconductivity, but we argue that it decreases in region \textbf{A} because $\Delta$$_{node}$ increases, not because T$_c$ decreases.

\begin{figure} [h]
\includegraphics [type=eps,ext=.eps,read=.eps,clip, width=3.5 in]{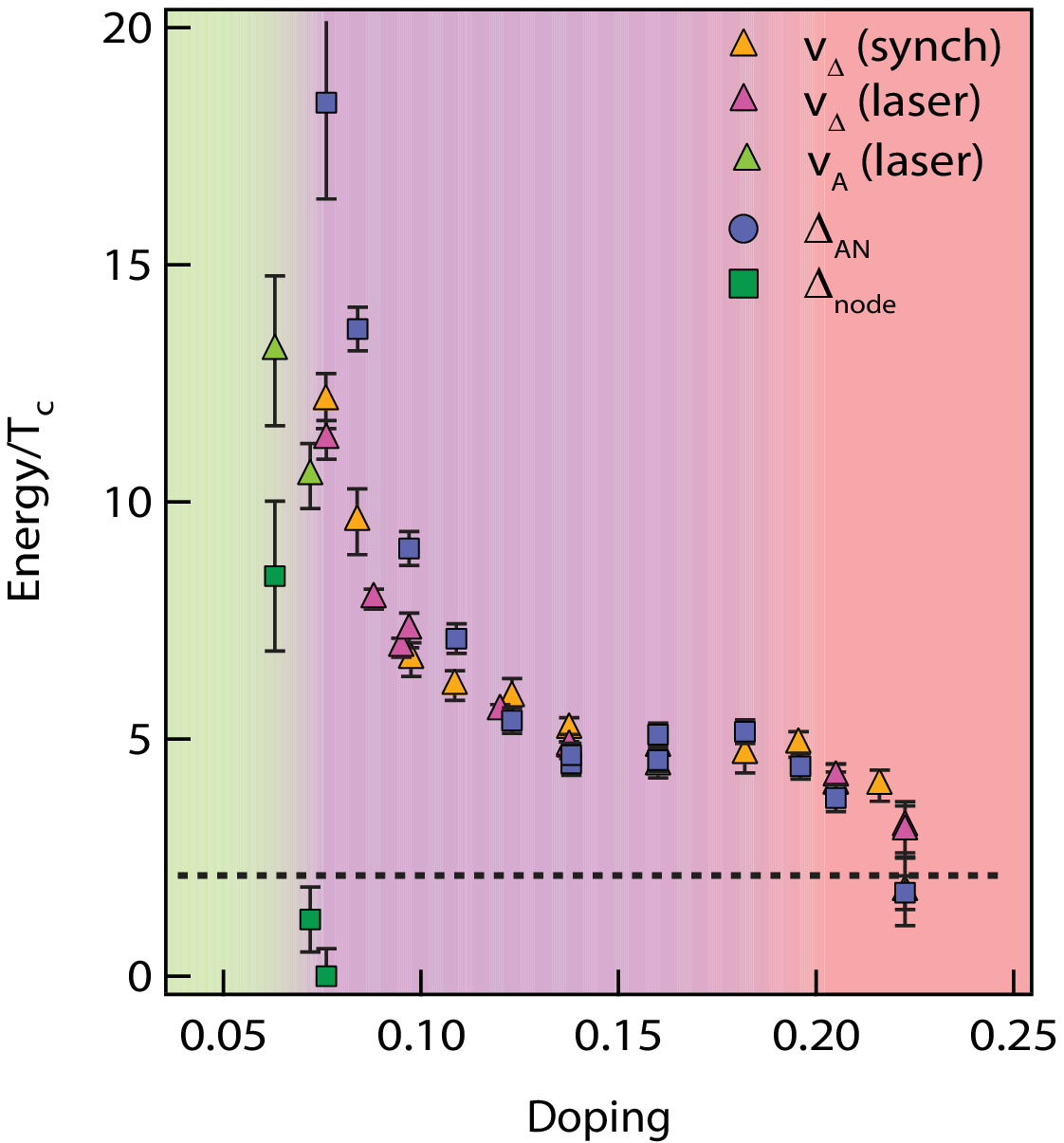}
\centering
\caption{\label{Fig 4_b:  Scaling with Tc} Fig. 2(d) of the manuscript, including also $\Delta$$_{AN}$, with energies scaled with T$_c$.  Horizontal dashed line denotes the \textit{d}-wave BCS ratio $\Delta$$/$T$_c$$=$2.14 }
\end{figure}

Fig. S\ref{Fig 4_b:  Scaling with Tc} shows the low-temperature energy scales plotted in Fig. 2(d) of the manuscript, scaled by T$_c$.

\section{Evolution from region B to region A}
\begin{figure*} [h]
\includegraphics [type=eps,ext=.eps,read=.eps,clip, width=5.5 in]{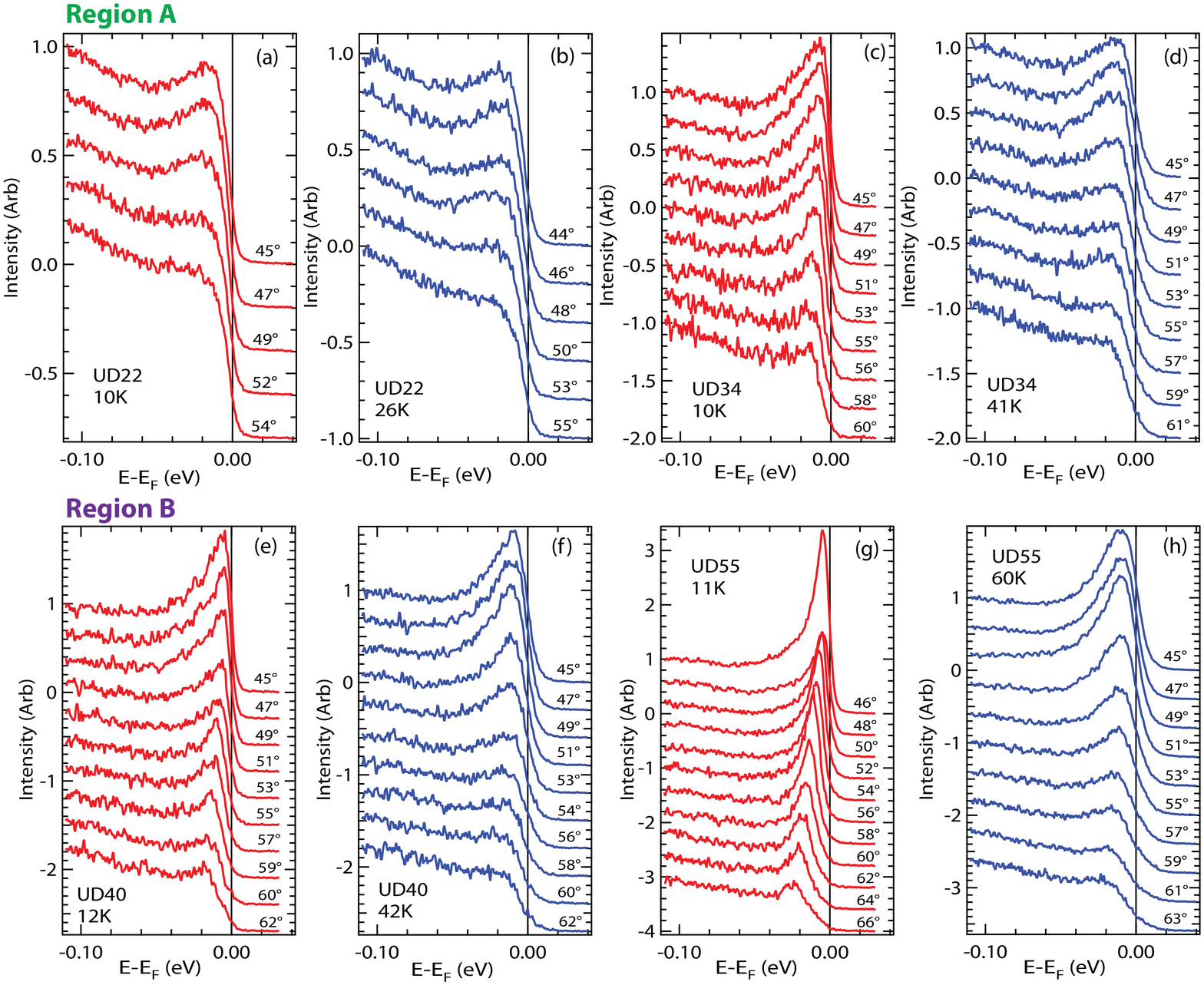}
\centering
\caption{\label{Fig 3b:  Raw EDCs region A and B} Raw EDCs at k$_f$ in region \textbf{A} and the most underdoped samples of region \textbf{B}, at low temperature (red) and above T$_c$ (blue).  EDCs are normalized to have equal intensity at 110meV. (a)-(d) Region \textbf{A}. Successive EDCs away from the node are shifted down by 0.2 (UD22) and 0.25 (UD34) in arbitrary units. (e)-(h) Region \textbf{B}. Successive EDCs away from the node are shifted down by 0.3 (UD40) and 0.4 (UD55) in arbitrary units. Sometimes different angles are sampled at different temperatures because of slight sample shifting.  Angles farther away from the node are shown for larger dopings because the quasiparticle weight increases with doping making peaks in off-nodal spectra increasingly more pronounced.}
\end{figure*}

Fig. S\ref{Fig 3b:  Raw EDCs region A and B} shows EDCs at k$_F$ below and above T$_c$ for samples in region \textbf{A} and \textbf{B}.  We point out several features.  First, the EDC at the nodal momentum in UD22 and UD34 exhibits a finite density of states at E$_F$.  Some of this is intrinsic to ARPES experiments, arising from scattered electron which have lost their momentum information.  The remainder may reflect a spatially and time varying phenomenon \cite{Haug:SDW_YBCO}, of which ARPES sees an average because of the large spot size and the time duration of data acquisition. Second, EDC widths at a given momentum show a smooth evolution from the most underdoped portion of region \textbf{B} into region \textbf{A}, indicating that samples in region \textbf{A} are not substantially more disordered than those in the underdoped part of region \textbf{B}.  We point out similar behavior in CCOC where low-energy peaks are observed in gapped spectra for p$=$0.10 \cite{KM_shen:fully_gapped}.

\begin{figure} [h]
\includegraphics [type=eps,ext=.eps,read=.eps,clip, width=2 in]{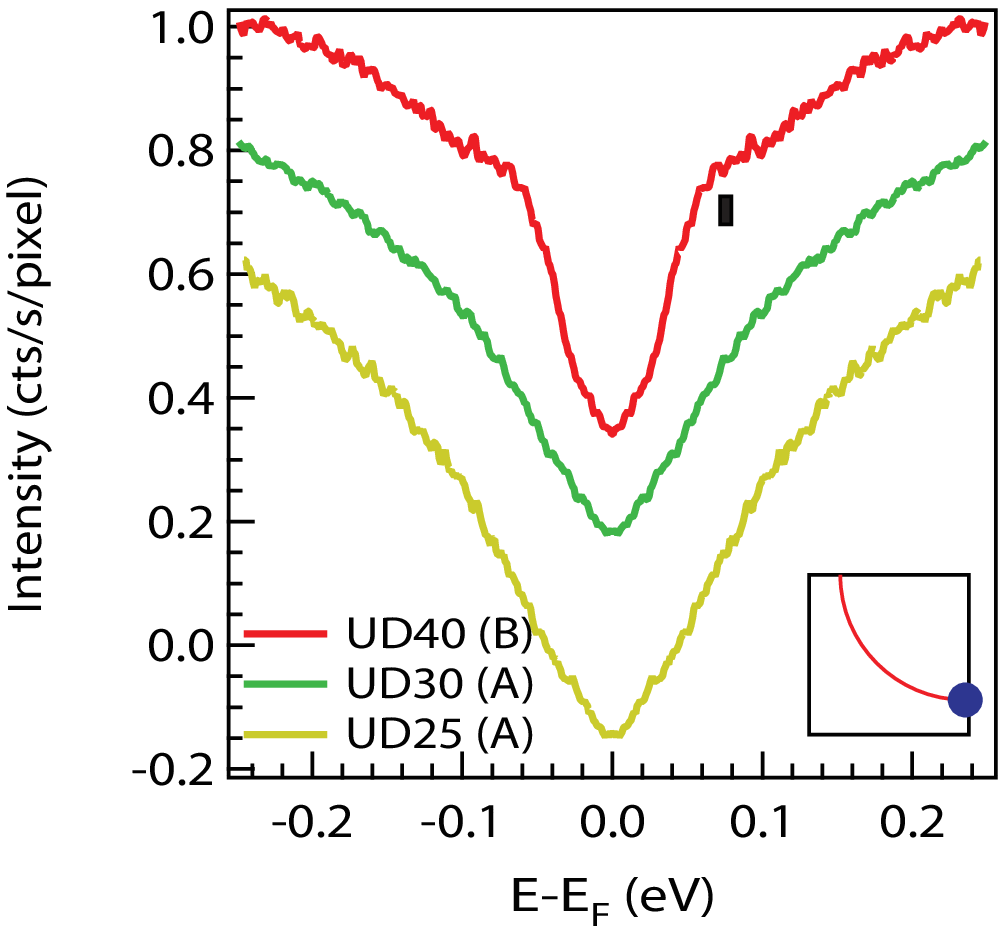}
\centering
\caption{\label{Fig 3d:  AN symmetrized spectra UD40 UD30 UD25} Antinodal symmetrized EDCs for dopings in region \textbf{A} (UD30, UD25) and region \textbf{B} (UD40).  Cuts taken parallel to $\Gamma$Y at 10K with 19eV photons in the second Brillouin zone. While UD40 shows remnants of quasiparticles at antinode below T$_c$, antinodal spectra for region \textbf{A} samples are featureless.}
\end{figure}

EDCs at the antinode (Fig. S\ref{Fig 3d:  AN symmetrized spectra UD40 UD30 UD25}) exhibit a change going from region \textbf{A} to region \textbf{B} at 10K.  While the latter shows a remnant of a quasiparticle peak, reflecting a gradual suppression of this feature with underdoping (see Fig. S\ref{Fig 4:  Comparison to kiyo}(c)), the former exhibits featureless antinodal spectra.  It is intriguing that antinodal quasiparticles are lost at the onset of region \textbf{A}.  We cannot dismiss the possibility that this is a matrix element or disorder effect, but it is also possible that this loss is intrinsic, perhaps arising from a change in FS topology.

\begin{figure} [h]
\includegraphics [type=eps,ext=.eps,read=.eps,clip, width=3.5 in]{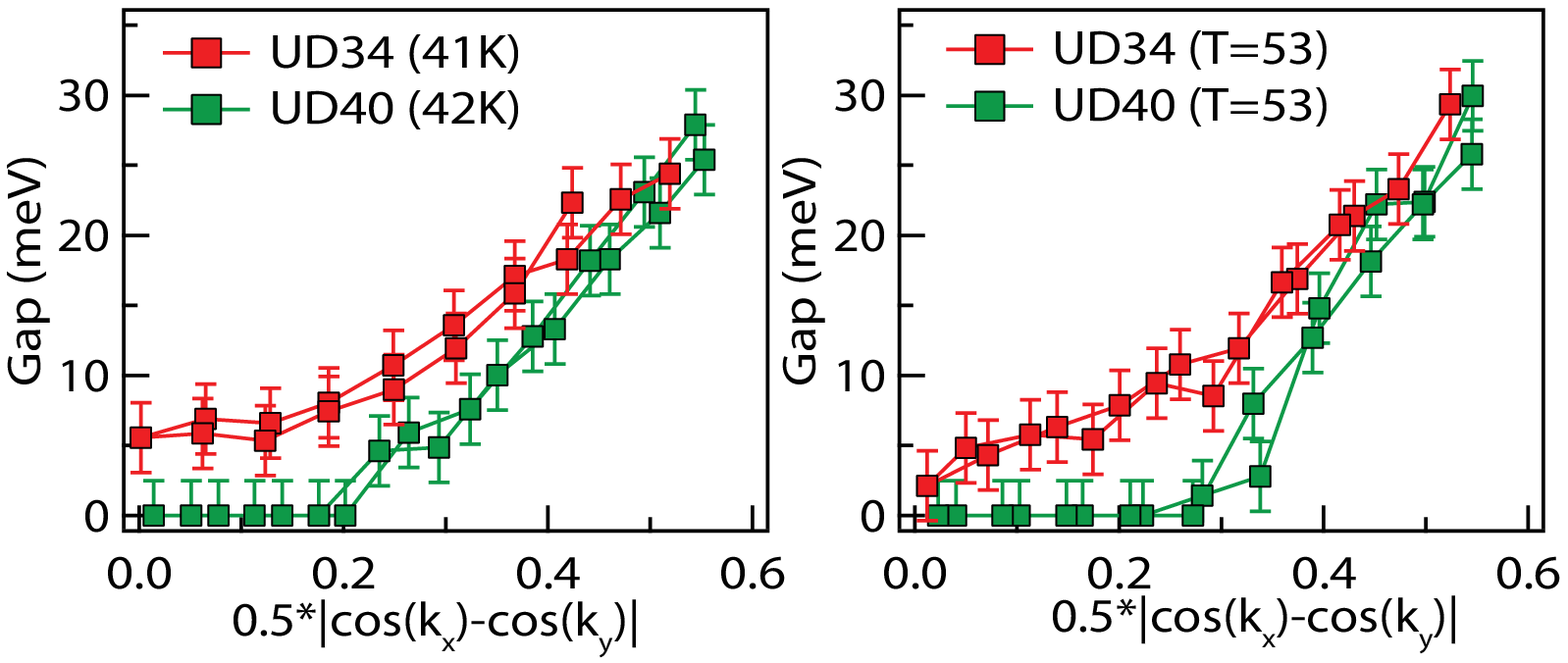}
\centering
\caption{\label{Fig 3c:  Gaps above Tc for UD34 and UD40} Gaps above T$_c$ for UD34 (p$\approx$0.072, region \textbf{A}) and UD40 (p$\approx$0.076, region \textbf{B}) at comparable temperatures.}
\end{figure}

Fig. S\ref{Fig 3c:  Gaps above Tc for UD34 and UD40} compares UD34 (region \textbf{A}) and UD40 (region \textbf{B}) gaps at similar temperatures above T$_c$.  While there is a small change in doping and T$_c$ between the two samples, the gap functions above T$_c$ are qualitatively different, with the former exhibiting a FS which is gapped at every momentum and the latter exhibiting characteristic pseudogap phenomenology with a Fermi arc.  At intermediate momenta, gaps are comparable.  This provides additional hints that the fully gapped FS in region \textbf{A} may be distinct from the pseudogap, with the pseudogap likely also persisting, though neutron scattering indicates that it may be weakened \cite{Haug:SDW_YBCO}.

\section{Fluctuating superconductivity}
\begin{figure} [h]
\includegraphics [type=eps,ext=.eps,read=.eps,clip, width=5 in]{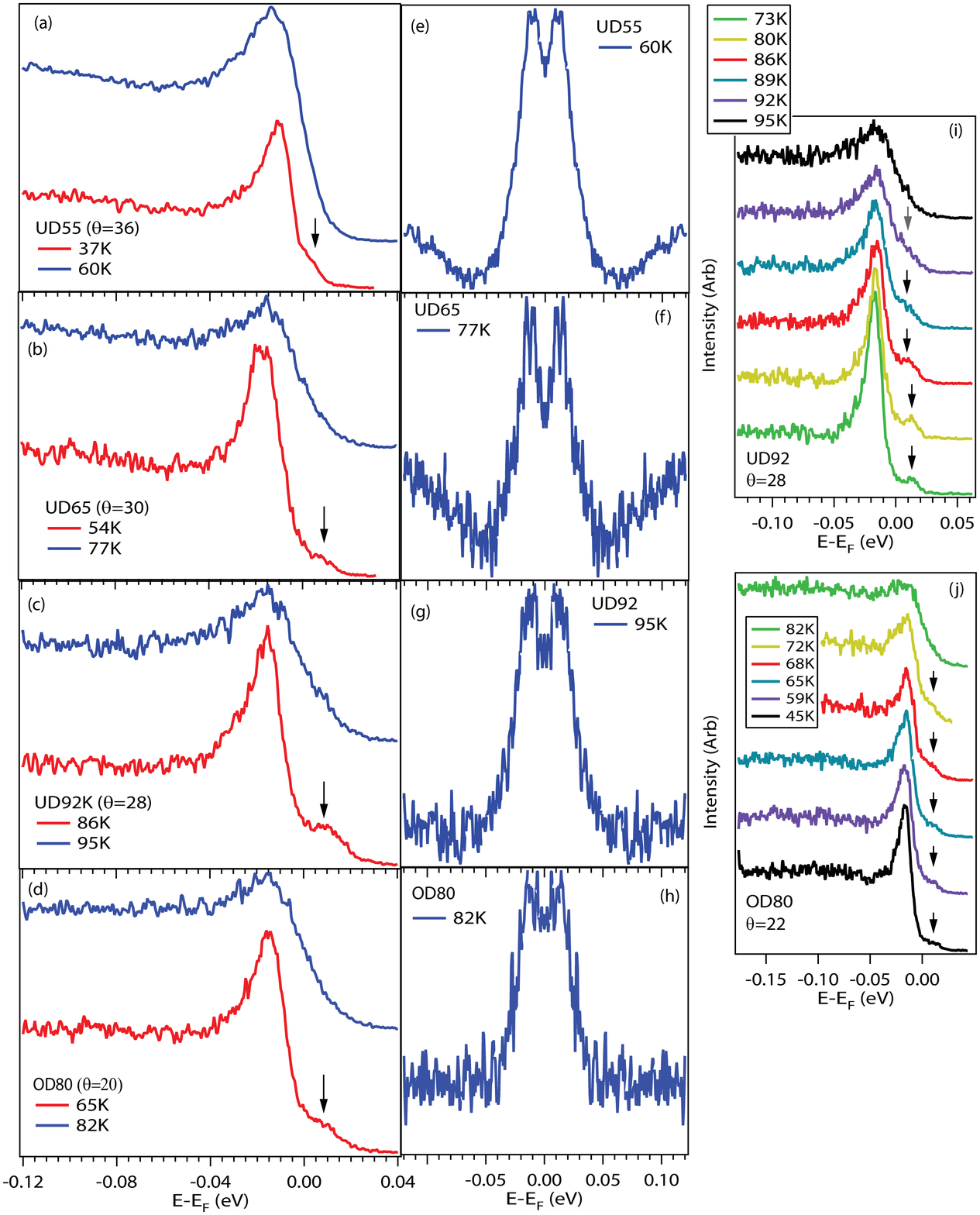}
\centering
\caption{\label{Fig 5:  Disappearance of upper Bogoliubov peaks above Tc} Disappearance of upper Bogoliubov peak above T$_c$.  EDCs at k$_F$, cut chosen to be in the gapped region for T slightly higher than T$_c$.  (a)-(d) EDCs at k$_F$ below (red) and above (blue) T$_c$ for UD55, UD65, UD92, and OD80.  Upper Bogoliubov peak is marked by arrow in T$<$T$_c$ data, but is not visible T$>$T$_c$. (e)-(h) Symmeterized EDCs for T$>$T$_c$, showing that spectra are still gapped at these momenta.  (i)-(j) Temperature dependence of EDC at k$_F$ for UD92 and OD80.  Arrows mark upper Bogoliubov peaks, which disappear across T$_c$.  FS angle $\theta$ defined in Fig. S\ref{Fig 1: EDCs at low temperature}.  }
\end{figure}

Evidence of superconducting fluctuations above T$_c$ has been reported by a number of techniques, some reporting a very large onset temperature \cite{Wang:Nernst,Li:DiamagnetismFluct} and other yielding an onset close to T$_c$ \cite{Orenstein:SC_fluct_Bi2212,Barisic:Microwave_SC_fluct_Hg1201}. In Fig. 4(c) of the manuscript,  we see a single spectral feature above T$_c$ whose phenomenology appears more consistent with the pseudogap, so we argue that while other experimental techniques can directly observe superconducting fluctuations above T$_c$, these features provide only a minority contribution to the spectral intensity seen by ARPES.  The first indicator of this is the disappearance of upper Bogoliubov peaks above T$_c$, as shown in Fig. S\ref{Fig 5:  Disappearance of upper Bogoliubov peaks above Tc}.  A superconducting gap of magnitude $\Delta$$_{SC}$ opens symmetrically at k$_F$, and an EDC at k$_F$ would have peaks at both $\omega$=$+$$\Delta$$_{SC}$ and $\omega$=$-$$\Delta$$_{SC}$ in the absence of a Fermi-Dirac cutoff.  At higher temperature, there is a small thermal population of states above E$_F$, and the enhanced photon flux of laser ARPES allows us to collect data with sufficient statistics to discern these upper Bogoliubov peaks.  The presence of the upper Bogoliubov peak is the clearest signature of superconductivity seen by ARPES in the cuprates, because much of the FS remains gapped above T$_c$ (the pseudogap) so a gap by itself does not signal superconductivity. The upper Bogoliubov peak is less pronounced in more underdoped samples, because the T$_c$ is lower, and the quasiparticle intensity tends to decrease with underdoping. EDCs at k$_F$ are shown below and above T$_c$ in Fig. S\ref{Fig 5:  Disappearance of upper Bogoliubov peaks above Tc} for four samples, and the peak/shoulder feature attributed to the upper Bogoliubov quasiparticle is marked by an arrow and shown to be absent above T$_c$.  A finer sampling of temperatures for OD80 and UD92 (Fig. S\ref{Fig 5:  Disappearance of upper Bogoliubov peaks above Tc}(e)-(f)) further illustrates the difference between superconducting spectra and non-superconducting spectra.  Notably, these data appear outside of the arc region of the pseudogap phase, defined as momenta where symmetrized EDCs imply zero gap, so if an upper Bogoliubov peak is present above T$_c$, we should be able to observe it at those momenta.  The second indicator that the gaps in Fig. 4(c) of the manuscript are of primarily pseudogap character is that they follow the well-established doping dependence of T* rather than the doping-independence of the superconducting gap in region \textbf{B}.

\section{Measuring T*}
\begin{figure} [h]
\includegraphics [type=eps,ext=.eps,read=.eps,clip, width=5.5 in]{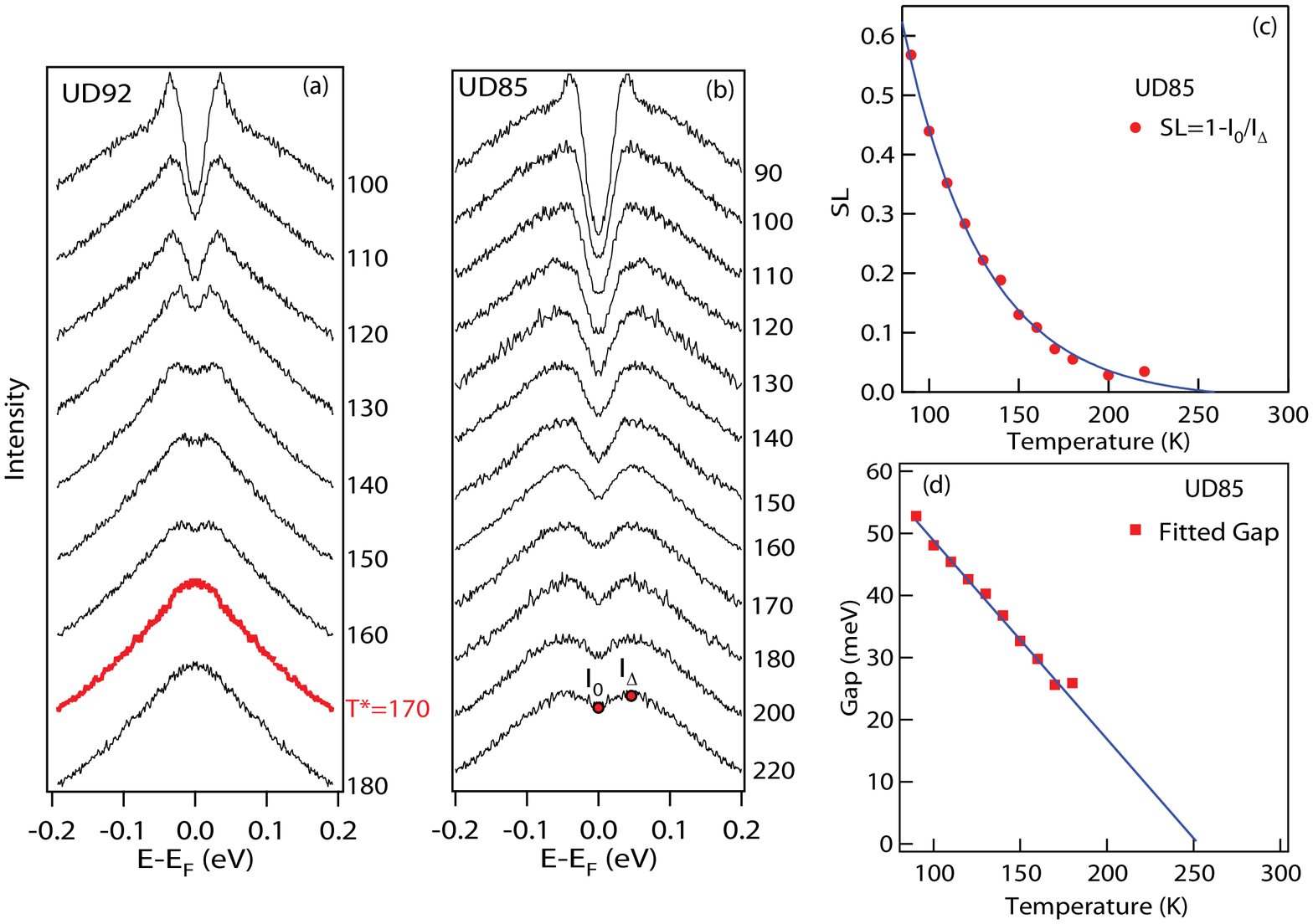}
\centering
\caption{\label{Fig 6:  T* from ARPES} Extracting T* from ARPES data. (a) UD92.  Symmetrized EDCs at k$_F$ T$>$T$_c$.  T* highlighted in red, defined as temperature when symmetrized EDCs show a single peak at E$_F$. (b) UD85, symmetrized EDCs at k$_F$.  (c)-(d) T* determined from extrapolating spectral loss function (SL)\cite{Kanigel:FermiArc} or fitted gap \cite{Symmetrization_Norman_model}.  Because antinodal spectra are considerably broader above T$_c$, an additional lifetime term is included in the fitting, as discussed in Ref. \cite{Symmetrization_Norman_model}. }
\end{figure}

In Fig. 4(f) of the manuscript we show T* from ARPES, STS, and SIS tunneling experiments together, because these are comparable techniques where T* is determined by a suppression of antinodal density of states at E$_F$.   If T* is sufficiently low to be accessible by ARPES, we use a standard definition \cite{Kondo:Fluct_vs_pseudogap,Kanigel:FermiArc}, defining T* as the temperature when symmetrized antinodal EDCs at k$_F$ exhibit a single peak at E$_F$, as shown in Fig. S\ref{Fig 6:  T* from ARPES} (a).  For more underdoped samples, T* is not reliably accessible by ARPES, because oxygen can become mobile above T$\approx$200K changing the doping near the surface during the course of an experiment.  In those cases, T* is determined by extrapolating parameters measured in the pseudogap state at lower temperature, such as the spectral loss function \cite{Kanigel:FermiArc} or the fitted gap \cite{Symmetrization_Norman_model}, as shown in Fig. S\ref{Fig 6:  T* from ARPES}(b)-(d).

\begin{figure} [h]
\includegraphics [type=eps,ext=.eps,read=.eps,clip, width=6 in]{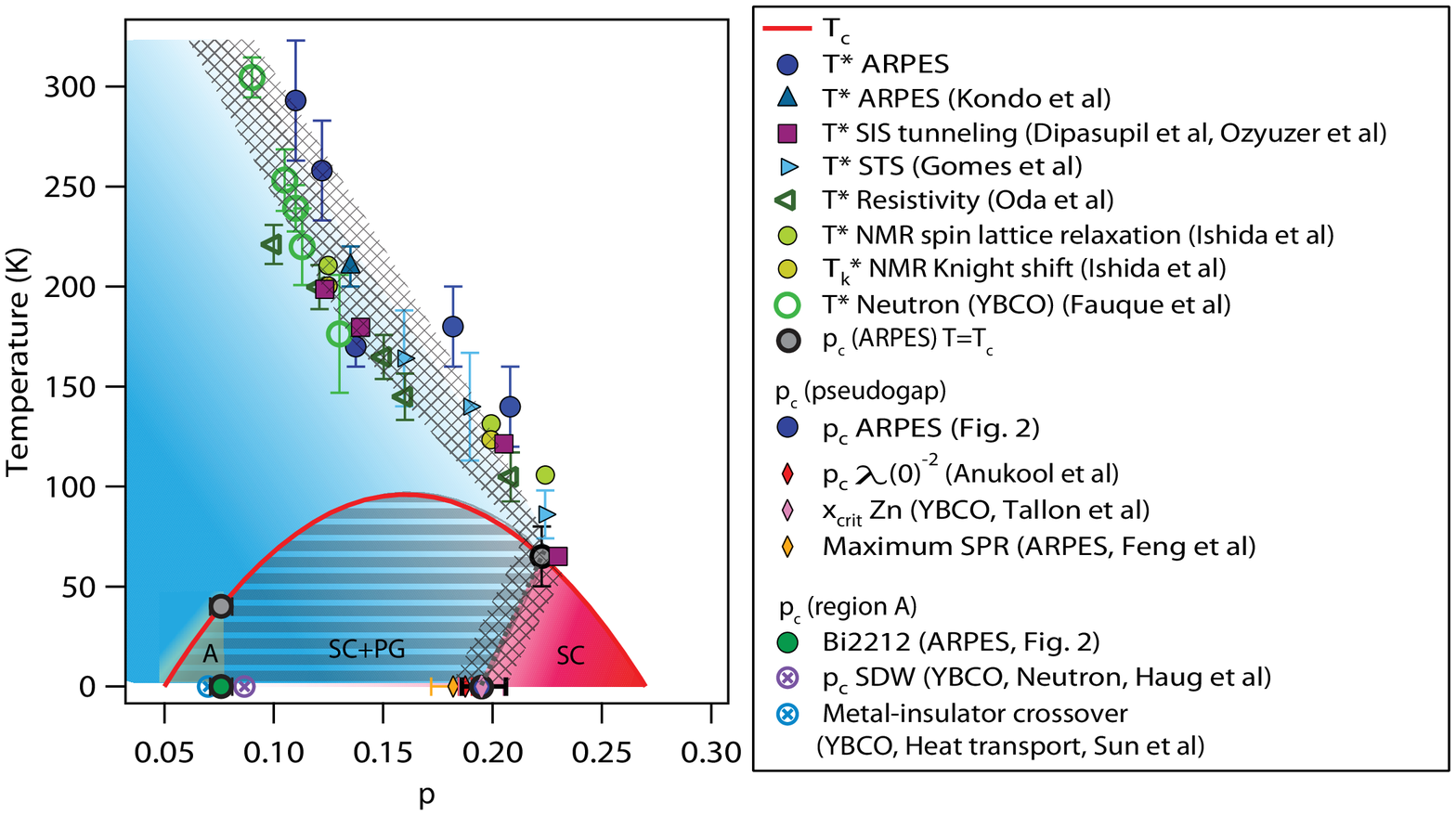}
\centering
\caption{\label{Fig 7:  T* from other experiments} Fig. 4 (f) from the manuscript and references therein, shown with T* data from other experiments: Resistivity \cite{Oda:PseudogapResistivity}, NMR \cite{Ishida:NMR_T_star_Bi2212}, and Neutron scattering on YBCO \cite{Fauque:NeutronYBCO}.  Low temperature measurements of pseudogap critical doping are indicated \cite{Anukool:LondonPenetration2009,Tallon:SF_density_cuprate_new_paradigm,Feng:SPR}.  Also shown are zero-field determinations of crossover (\cite{Sun:MI_crossover_YBCO}) or critical (\cite{Haug:SDW_YBCO}) dopings measured in YBCO, together with the onset doping of region \textbf{A} from ARPES. }
\end{figure}

T* from other experimental techniques (in-plane resistivity, NMR, neutron scattering) are shown in Fig. S\ref{Fig 7:  T* from other experiments}   \cite{Oda:PseudogapResistivity,Ishida:NMR_T_star_Bi2212,Fauque:NeutronYBCO}.  Neutron scattering data is shown for YBCO in Fig. S\ref{Fig 7:  T* from other experiments}, because data on Bi-2212 is currently not published.  Though a number of experiments support a critical point of the pseudogap at p$=$0.19 \cite{Anukool:LondonPenetration2009,Tallon:SF_density_cuprate_new_paradigm}, there are data from a number of experiments (transport, NMR, ARPES, tunneling) indicating a pseudogap persisting above T$_c$ for p$>$0.19, as seen in Fig. S\ref{Fig 7:  T* from other experiments}; this is reconciled in the manuscript via evidence of phase competition between superconductivity and the pseudogap.  Fig. S\ref{Fig 7:  T* from other experiments} also plots data from experiments which directly observe a pronounced change in ground state superconducting properties at p$=$0.19, consistent with the critical point of the pseudogap: superfluid density \cite{Anukool:LondonPenetration2009,Tallon:SF_density_cuprate_new_paradigm}, superconducting peak ratio \cite{Feng:SPR}, and Cu-site doping required to destroy superconductivity\cite{Tallon:SF_density_cuprate_new_paradigm}.  As discussed in the main text, a number of experiments on YBCO report an emergent phase at the underdoped edge of the superconducting dome, perhaps related to phase region \textbf{A} observed in Bi-2212.  For comparison to ARPES data only zero magnetic field or low magnetic field results are shown in Fig. S\ref{Fig 7:  T* from other experiments} \cite{Haug:SDW_YBCO,Sun:MI_crossover_YBCO}, though we note that high field experiments yield similar critical dopings \cite{Sebastian:Diverging_effective_mass_QO,LeBoeuf:LifshitzCriticalPoint_YBCO}.

\clearpage

\makeatletter
\makeatletter \renewcommand{\fnum@figure}
{\figurename~S\thefigure}
\makeatother

